\documentclass[conference]{IEEEtran}

\ifCLASSINFOpdf
  \usepackage[pdftex]{graphicx}
  \graphicspath{{./images/}}
  \DeclareGraphicsExtensions{.pdf}
\else
\fi

\newcommand{\TheName}{\textsc{DexLego}}

\usepackage[normalem]{ulem}
\usepackage{verbatim}
\usepackage{multirow}
\usepackage{multicol}
\PassOptionsToPackage{hyphens}{url}
\usepackage{hyperref}
\usepackage{url}

\usepackage{amssymb}
\usepackage{pifont}
\usepackage{subcaption}
\usepackage{etoolbox}
\usepackage{booktabs,fixltx2e}
\usepackage[flushleft]{threeparttable}
\usepackage{listings}
\usepackage{color}
\usepackage{caption}
\usepackage{algorithm}
\usepackage{algpseudocode}
\usepackage{pgfplots}
\usepackage{hyperref}
\usepackage{amsmath}
\usepackage{afterpage}
\usepackage{pdflscape}
\usepackage{makecell}
\usepackage[noadjust]{cite}
\usetikzlibrary{patterns}
\usetikzlibrary{positioning,calc}

\pgfplotsset{compat=1.7}

\makeatletter
\renewcommand{\ALG@beginalgorithmic}{\footnotesize}
\makeatother
\algrenewcommand\alglinenumber[1]{\tiny #1:}

\usepackage{pifont}
\usepackage{balance}
\usepackage{csquotes}

\definecolor{dkgreen}{rgb}{0,0.6,0}
\definecolor{gray}{rgb}{0.5,0.5,0.5}
\definecolor{mauve}{rgb}{0.58,0,0.82}

\pgfplotsset{
    error bars with mapped color/.style={
        disabledatascaling,
        visualization depends on=\thisrow{#1} \as \error,
        visualization depends on=\thisrow{y} \as \y,
        scatter/@pre marker code/.append style={
            /pgfplots/error bars/.cd,
            error mark options={draw=mapped color},
            error mark=|,
            draw error bar={(0,0)}{(0,\error*100)}, 
            draw error bar={(0,0)}{(0,-\error*100)} 
        },
        scatter/@post marker code/.append code={}
    }
}

\hypersetup{hidelinks,colorlinks=true,citecolor=blue,urlcolor=black,linkcolor=blue}

\renewcommand{\lstlistingname}{Code}

\begin{document}
%
\title{DexLego: Reassembleable Bytecode Extraction for Aiding Static Analysis}
\author{
	\IEEEauthorblockN{Zhenyu Ning and Fengwei Zhang}
	\IEEEauthorblockA{COMPASS Lab}	
    \IEEEauthorblockA{Wayne State University}	
    \IEEEauthorblockA{\{zhenyu.ning, fengwei\}@wayne.edu}
}

\maketitle

\begin{abstract}


The scale of Android applications in the market is growing rapidly. 
To efficiently detect the malicious behavior in these applications, an array of static 
analysis tools are proposed. However, static analysis tools suffer from code hiding techniques like
packing, dynamic loading, self modifying, and reflection.
In this paper, we thus present \TheName{}, a novel system that performs a reassembleable bytecode extraction for aiding static analysis tools 
to reveal the malicious behavior of Android 
applications. \TheName{} leverages just-in-time 
collection to extract data and bytecode from an application at runtime, and reassembles them to a new 
Dalvik Executable (DEX) file offline. The experiments on DroidBench and real-world applications show
that \TheName{} correctly reconstructs the behavior of an application in
the reassembled DEX file, and significantly improves analysis result of the existing static analysis systems.
\end{abstract}

\section{Introduction}
\label{sec:intro}

With the rapid proliferation 
of malware attacks on mobile devices, understanding their malicious behavior 
plays a critical role in crafting effective defense. Static 
analysis tools are used to analyze malware and investigate their
malicious activities~\cite{flowdroid:pldi14,horndroid:eurosp16,droidsafe:ndss15,iccta:icse15,amandroid:ccs14}. 
However, malware writers can hide the malicious behavior by using an array of obfuscation techniques. 
The annual report from AVL team~\cite{report:avl15} 
shows that the number of Android packed applications has increased
more than nine times, while about one third of them are packed malware.
Typically, static analysis tools identify the malicious behavior of an application by investigating
bytecode in Dalvik Executable (DEX) files.
The packing technology replaces the original DEX file with a shell DEX file and dynamically
releases the original DEX file at runtime. Additionally, the original DEX file is encrypted until its execution. While the free use of 
public packing platforms~\cite{ali:protector,baidu:protector,ijiami:protector,dex:protector,360:protector,
tencent:protector} provides a convenient and reliable protection for
applications, the challenge of facing packed malware is rising. Static analysis tools are
completely unarmed to the packed malware as they can only fetch the shell DEX file
but not the encrypted original DEX file. 

To address this problem, several unpacking systems are introduced recently~\cite{appspear:raid15,dexhunter:esorics15}. However, these systems are far from solving the problem completely. For instance, they assume
that there is a point when all original code is unpacked in memory (i.e., a clear boundary or transition between the packer's code and the original code). 
However, the malware writers can pack code with 
advanced techniques that interleave the packing and unpacking processes. Moreover, recent studies 
show that sophisticated adversaries, known as self-modifying malware~\cite{asac:bb,dabid:bh15}, can 
modify the bytecode and other contents in a DEX file at runtime.

To further understand the self-modifying malware, consider Code~\ref{lst:smc} as an example.
In Line $14$, the native method, \texttt{bytecodeTamp\-er}, modifies the bytecode 
of Lines $11$ and $13$. Note that the method \texttt{bytecodeTamper} is executed 
twice and performs different modifications to the two Lines during each iteration. 
There is a taint flow in Code~\ref{lst:smc}, but the state-of-the-art static analysis 
tools~\cite{flowdroid:pldi14,horndroid:eurosp16,droidsafe:ndss15,iccta:icse15, amandroid:ccs14} 
cannot detect it. Moreover, existing method-level unpacking systems~\cite{appspear:raid15,dexhunter:esorics15} 
are unable to reveal this taint flow because they cannot differentiate the actual 
executed code from the fake code (i.e., modified code like Lines $11$ and $13$ to hide taint flows), 
and we will 
discuss the details in Section~\ref{ssec:bcd}.

Unlike the static analysis tools, the dynamic analysis tools~\cite{taintdroid:osdi10, taintart:ccs16, copperdroid:ndss15,
droidscope:usenix12, vetdroid:ccs13} do not suffer from packing techniques.
However, they have their own drawbacks. The automatic dynamic taint flow 
analysis tools~\cite{taintdroid:osdi10, taintart:ccs16, vetdroid:ccs13} cannot handle implicit
taint flows while static analysis tools~\cite{iccta:icse15,amandroid:ccs14} can solve them. Moreover,
the huge performance overhead makes it difficult to implement a complicated analysis mechanism, so
there is a trade-off between the accuracy and performance. Meantime, the code coverage
problem also threatens the accuracy of the dynamic analysis tools~\cite{copperdroid:ndss15, droidscope:usenix12, vetdroid:ccs13}.

In this paper, we present \TheName{}, a novel program transformation system that reveals the hidden code in 
Android applications
to analyzable pattern via instruction-level extracting and reassembling.
\TheName{} collects
bytecode and data when they are executed and accessed, and reassembles 
the collected result into a valid
DEX file for static analysis tools. 
Since we extract all executed
instructions, our system is able to uncover the malicious
behavior of the packed applications or malware with self-modifying code.
One of the key challenges in \TheName{} is to reassemble the instructions into a valid and accurate DEX file. 
Hence, we design a novel reassembling approach to construct the entire executed control
flows including self-modifying code. Additionally, we implement
the first prototype of force execution on Android and use it as our
code coverage improvement module. 

Moreover, our system helps static analysis tools improve the analysis
accuracy on reflection-involved samples. The Java reflection obscures the control flows
of the application by replacing the direct function call or field access with a call to the
reflection library functions which take the name string of the function or field as parameter.
Previous reflection solutions~\cite{barros:ase15} 
and static analysis tools~\cite{flowdroid:pldi14,horndroid:eurosp16,droidsafe:ndss15} on Android
assume that the name strings of the reflectively invoked method and its declaring class are reachable. 
However, the name string can be encrypted in some cases~\cite{harvester:ndss16} and the advanced 
malware could even use reflective method calls without involving any string parameter~\cite{droidbench:ecssse}. 
A solution on traditional Java platform~\cite{tamiflex:icse11} requires load-time instrumentation 
which is not supported in Android~\cite{flowdroid:pldi14}.
Thus, \TheName{} implements a similar idea in Android and replaces the reflective call with direct call.

\setlength{\textfloatsep}{0pt}
\begin{lstlisting}[caption={An Example of Self-Modifying Code.},float=tp, label={lst:smc}]
package com.test;

public class Main extends Activity {
  private static final String PHONE = "800-123-456";
  protected void onCreate(Bundle savedInstanceState) {
    // ...
    advancedLeak();
  }
  
  public void advancedLeak() {
    String a = getSensitiveData(); // source
    for (int i = 0; i < 2; ++i) {
      normal(a);
      bytecodeTamper(i);
    }
  }
 
  public void normal(String param) {
    // do something normal 
  }
  
  public void sink(String param) {
    // send param through text message.
	SmsManager.getDefault().sendTextMessage(PHONE, null, param, null, null); // sink
  }
  
  /* While i = 0:
   *   modify Line 11 to String a = "non-sensitive data"
   *   modify Line 13 to sink(a)
   * While i = 1:
   *   modify Line 11 to String a = getSensitiveData()
   *   modify Line 13 to normal(a) */
  public void native bytecodeTamper(int i);
}
\end{lstlisting}

We evaluate \TheName{} on real-world packed applications and DroidBench~\cite{droidbench:ecssse}. 
The evaluation result shows \TheName{} successfully unpack and reconstruct the behavior of the applications. 
The F-measures (i.e., analysis accuracy) of FlowDroid~\cite{flowdroid:pldi14}, DroidSafe~\cite{droidsafe:ndss15}, 
and HornDroid~\cite{horndroid:eurosp16} on DroidBench increase $33.3\%$, $31.1\%$, 
and $23.6\%$, respectively. Moreover, 
static analysis tools with the help of \TheName{} provide a better accuracy than existing
dynamic analysis systems TaindDroid~\cite{taintdroid:osdi10} and TaindART~\cite{taintart:ccs16}.
The code coverage experiments on open source samples from F-Droid~\cite{fdroid:fdroid}
show that our force execution module helps to improve the coverage of dynamic analysis
and increases the coverage of state-of-the-art fuzzing tool, Sapienz~\cite{sapienz:issta16}, from $32\%$ to $82\%$.
The main contributions of this work include:

\begin{itemize}

\item We present \TheName{}, a novel system that automatically transforms the hidden code in 
the Android applications
to analyzable pattern. Our novel approach leverages tree structures to collect data/bytecode at runtime, and reassemble collected information back to DEX files, which makes
the hidden code including packed or self-modifying one analyzable for current static analysis tools. 
To the best of our knowledge, this is
the first system to reassemble the instruction-level tracing result of Java bytecode back to an executable file, and we consider this is the \textbf{key contribution} of this work. 

\item \TheName{} mitigates the inaccuracy of static analysis tools on the 
reflection-involved samples by transforming the reflective method call to
direct call regardless how the adversary uses it; it also improves the code coverage of dynamic analysis via our force execution
module and. Moreover, \TheName{} can be
easily applied to Java application on x86 platforms and advances the traditional taint flow
analysis.

\item We implement a prototype of \TheName{} in Android Runtime and evaluate the system in a real 
Android device. 
The experiment result shows that \TheName{} successfully unpacks and reconstructs the hidden behavior of the 
real-world packed applications.
By testing our system with state-of-the-art static analysis
tools on DroidBench, we demonstrate that \TheName{} improves the F-Measures of static analysis tools 
by more than $23\%$.
Moreover, the comparison with existing dynamic analysis tools shows that \TheName{}-assisted 
approach provides a more accurate result. 

\item The source code of \TheName{} is publicly available at \texttt{goo.gl/jpRvqu}.
\end{itemize}

\section{Background}
\label{sec:background}

\subsection{Dalvik and Android Runtime}
\label{ssec:dalvik&art}

Dalvik is a special Java virtual machine running in the Android system. It is used to interpret
Android specified bytecode format since the first release of Android. To improve 
the performance, Google has introduced Just-In-Time (JIT) compilation
and Ahead-Of-Time (AOT) compilation since Android 2.2
and Android 4.4, respectively. The JIT compilation continually compiles frequently executed
bytecode slices into the machine code. As an upgrade, the AOT compilation compiles most bytecode in the application into 
the machine code during the installation. Dalvik equipped with AOT compilation is renamed to Android Runtime (ART).
Since Android 5.0, Dalvik has been completely replaced by ART. 

In both Dalvik and ART, the bytecode is organized in units of methods. The minimum code unit for JIT and AOT compilation
is a method, indicating that a single method cannot contain both bytecode and machine code.
Methods such as constructors and abstract methods require the bytecode interpreter even in ART. 
Moreover, a single method or the entire ART can be configured to 
run in the interpreter mode.

\subsection{Android Java Bytecode}

The Java bytecode in Android is chained by instructions. Each instruction contains 
an opcode and arguments related to the opcode. The opcodes are different from the ones in regular
Java bytecode and the bit-length of an instruction varies according to the opcode. 
In the interpreter, instructions are listed in an array of 16-bit ($2$ bytes) units. 
An instruction 
occupies at least one unit with a maximum number of units up to five. 
\section{System Overview}
\label{sec:architecture}

\setlength{\textfloatsep}{15pt}
\begin{figure}
\centering
\includegraphics[width=3.2in]{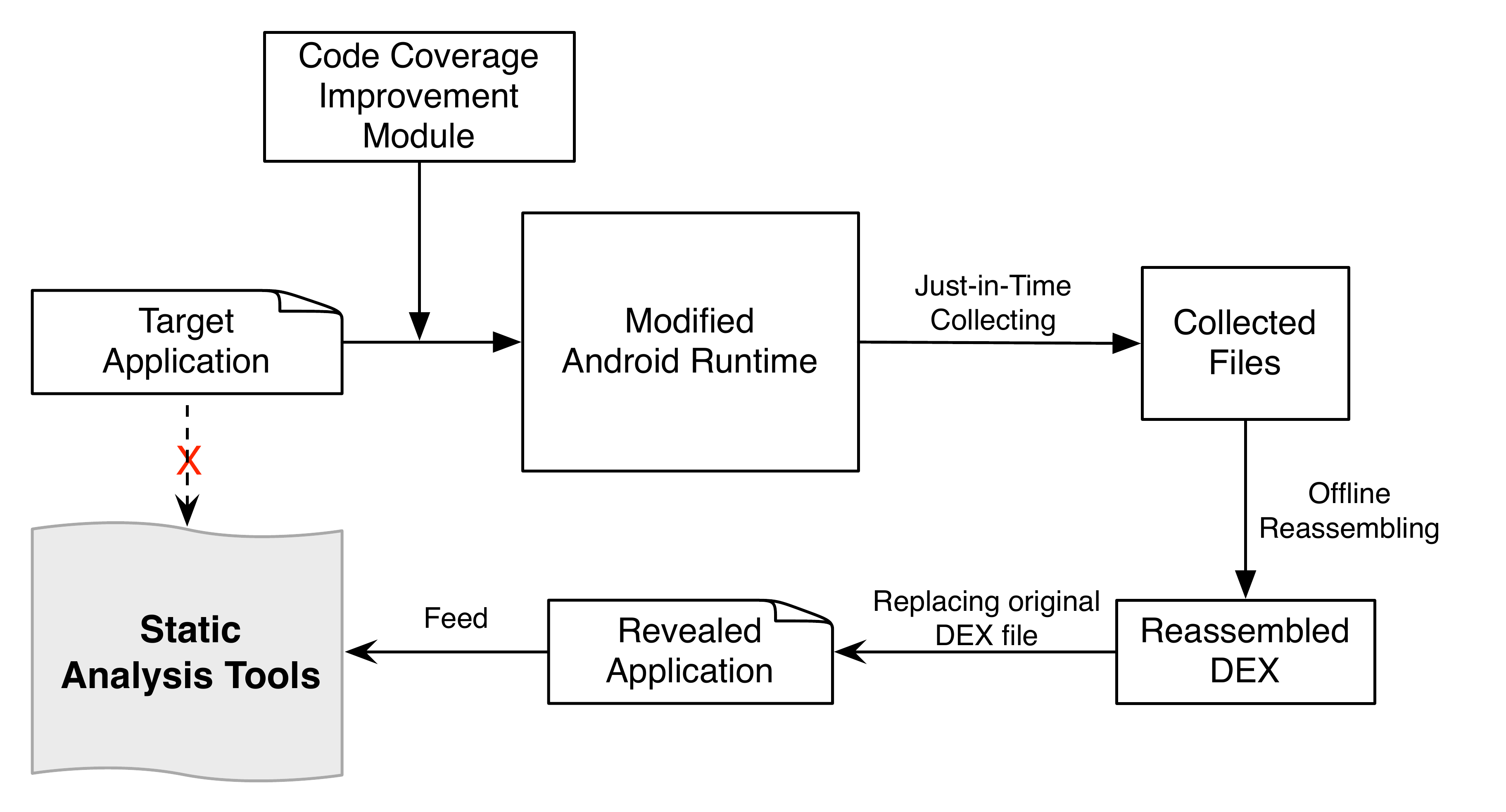}
\caption{Overview of \TheName{}.}
\label{fig:arch}
\vspace{-10pt}
\end{figure}

As Figure~\ref{fig:arch} shows, instead of directly feed the target application to static analysis
tools, we firstly execute the target application with \TheName{}. 
In executing, we use Just-in-Time (JIT) 
collection to extract data/instructions and output them
to files right before used by ART. In the meantime, we use a code coverage improvement module to increase the code coverage. Next, we reassemble the collected files to
a DEX file and use the reassembled DEX file to replace the one in the original APK. 
Finally, the new APK file is fed to the static analysis tools. The architecture of \TheName{}
contains three main components: 1) the collecting component that collects bytecode and data, 
2) the offline reassembling component that reassembles a new DEX file based on the collection result,
and 3) the code coverage improvement module that helps \TheName{} to achieve a high code
coverage. Next, we will discuss the three components respectively.

\subsection{Bytecode and Data Collection}
\label{ssec:abdc}

\begin{figure}
\centering
\includegraphics[width=2.8in]{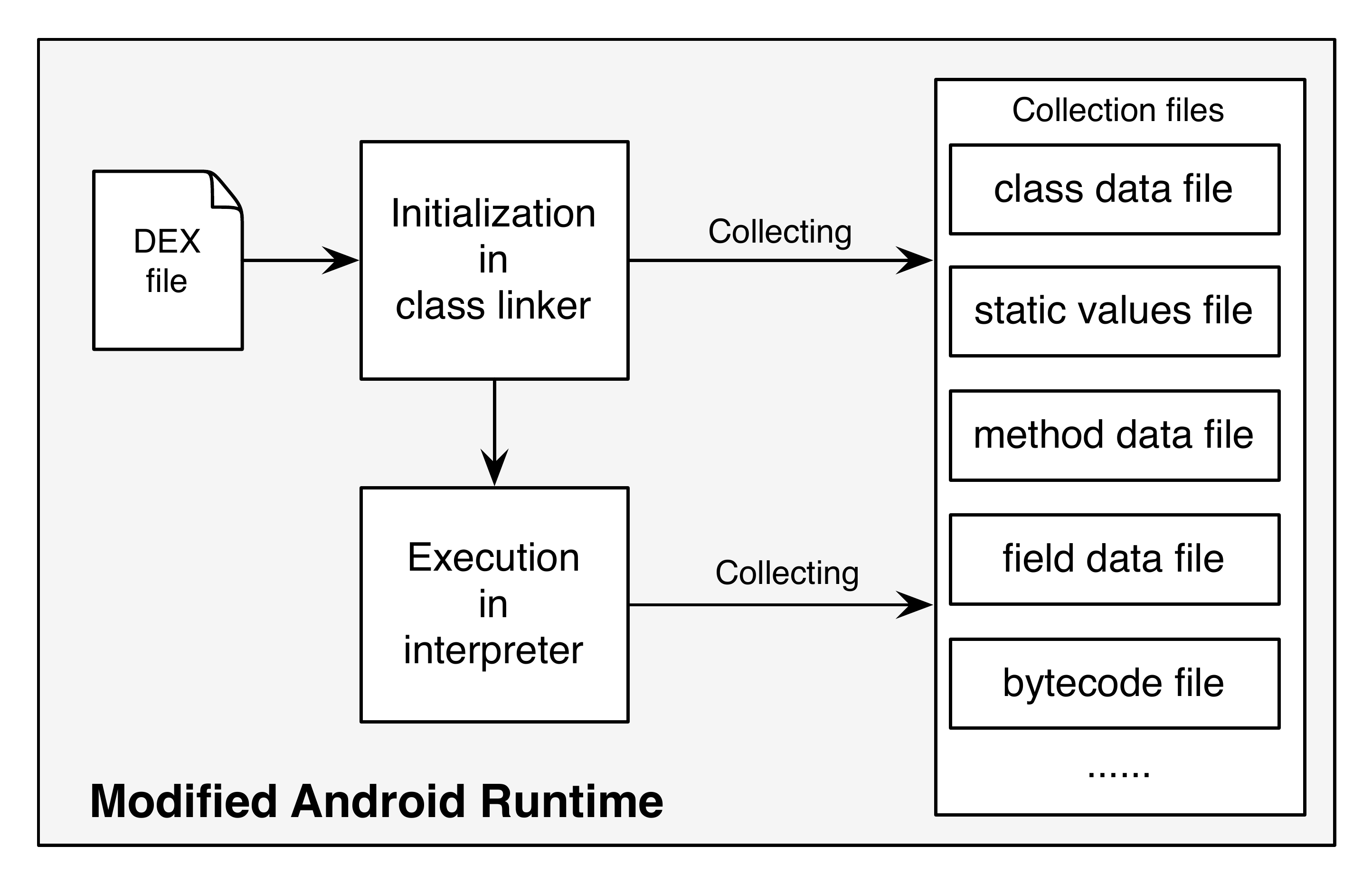}
\caption{Just-in-Time Collection.}
\label{fig:archextract}
\vspace{-10pt}
\end{figure}

Figure~\ref{fig:archextract} shows the JIT collection we used in \TheName{}.
During the execution of an application, ART firstly extracts the DEX file from the original APK file
and passes it to the class linker. The class linker then loads and initializes the classes in the DEX file,
and our JIT collection method collects the metadata of the class (e.g., super class) at this
point. Next, when a method is invoked, ART extracts its bytecode from the DEX file,
and leverages the interpreter to execute them.
The interpreter fetches the entire bytecode (organizing in a 16-bit array) of the method and executes
the bytecode instructions one by one. Thus, according to our JIT policy, we collect the 
executed instructions of the method and their related objects (e.g., \texttt{string})
via instruction-level extracting. Note that the execution of the code in the dynamic loaded DEX file also
follows the same flow.

\setlength{\textfloatsep}{15pt}

The state-of-the-art static analysis tools do not accept machine
code as their input. However, ART executes most methods based on the machine code, and
the translation from the machine code to the bytecode is a challenging task. 
To simplify the task, \TheName{} configures all methods in the application to be executed by the interpreter.

\subsection{DEX File Reassembling}
\label{ssec:reassembling}

After the collecting, all the output files are reassembled to a new DEX file offline following 
the format of a DEX file, and we replace the DEX file in the original APK file with the reassembled one. 
The modified APK file is finally fed to static analysis tools to study the malicious behavior. 

This reassembling is not trivial, and we consider this is the key contribution of this work. In the DEX file format, each method contains only 
one instruction array. However, due to different control flows (e.g., execution is led to different branches
of a branch statement) or self-modifying code, one method may 
contain different instruction arrays in the collection stage. To correctly combine the collected 
instructions, we thus design a tree model and a novel collecting and reassembling mechanism.
More details are discussed in Section~\ref{ssec:bcd} and Section~\ref{ssec:bcr}.

\subsection{Code Coverage Improvement Module}
\label{ssec:codecoverage}

To improve the code coverage of dynamic analysis systems, there already exists a series of tools
or theories like: 1) Input generators or fuzzing 
tools~\cite{guiripper:ase12,a3e:oopsla13,monkey:gg,puma:mobisys14,dynodroid:fse13}, 
2) Symbolic or concolic execution~\cite{anand:fse12,klee:osdi08, se:sigsoft12,harvester:ndss16,
intellidroid:ndss16,appintent:ccs13} based systems,
3) Force execution~\cite{iris:ccs15,jforce:www17,xforce:usenix14} based systems.
Our code coverage improvement module can be one of them or a combination of them.
Note that most of the systems mentioned in 1) and 2) are implemented in Android, and we 
can directly use them to conduct the execution of the target application with little 
engineering effort. However, to the best of our knowledge, the idea of force execution has not 
been applied on Android platform. Thus, we implement a prototype of force execution as a supplement of our
code coverage improvement module.

To use force execution in \TheName{}, we identify the Uncovered Conditional
Branches (UCB) and calculate the path to each UCB. By monitoring and
manipulating the branch instructions in the interpreter, we force the control
flow to go along the calculated path to reach each UCB.
\section{Design and Implementation}
\label{sec:implementation}

We implement \TheName{} in an LG Nexus 5X with Android 6.0.
Based on the Android Open Source Project~\cite{asop:gg} (AOSP), we build a customized system image 
and flash it into the device by leveraging a third-party recovery system~\cite{twrp:teamwin}. 

A DEX file consists of data structures that represent different data types used by the 
interpreter~\cite{dex:gg}. 
\TheName{} collects these data structures directly from memory while 
they are used by ART at the runtime.
Moreover, we leverage instruction-level tracing to collect executed instructions 
and reassemble them back to a \texttt{method} structure. 
In this section, we discuss 1) bytecode collection, 
2) bytecode reassembling, 3) data collection, and 4) DEX file reassembling separately.
The approaches to handle reflection and force execution are also discussed
in this section.

\subsection{Bytecode Collection}
\label{ssec:bcd}

In ART, after the instruction array of a method is passed to the interpreter, the interpreter 
executes the instructions one by one following the control flow indicated by them. To expose
the behavior of the method,  \TheName{} aims to collect all instructions executed in the
method. However, existing systems~\cite{appspear:raid15,dexhunter:esorics15} that use method-level collection cannot defend against dynamic bytecode modification, and the detailed limitation is described as below. 

\noindent\textbf{Inadequacy of Method-level Collection.} 
Consider Code~\ref{lst:smc}
as an example. While entering the method \texttt{advancedLeak}, the smali code~\footnote{
The smali code is a more readable format of the bytecode.}
of the method is represented by Code~\ref{lst:el}. After the first execution of the native 
method \texttt{bytecodeTamper}, the code of the method \texttt{advancedLeak} is modified to 
Code~\ref{lst:fe}. In Code~\ref{lst:fe}, the native method has modified the bytecode 
to hide the source (Lines $2$-$4$ are changed from Code~\ref{lst:el} to Code~\ref{lst:fe}), 
but the sensitive data is already stored in the register \texttt{v0}. 
During the second execution of the \texttt{for} loop, the sensitive data in the register \texttt{v0}
is leaked through the method \texttt{sink} (Lines $9$-$10$ in Code~\ref{lst:fe}). Then, the native 
method resumes the code back to Code~\ref{lst:el}. The instruction array of the method
\texttt{advancedLeak} in memory is either Code~\ref{lst:el} or
\ref{lst:fe} at any time point (e.g., before and after JNI code), which means
that the method-level collection (e.g., DexHunter~\cite{dexhunter:esorics15} and AppSpear~\cite{appspear:raid15}) can only
collect Code~\ref{lst:el} or \ref{lst:fe} even when multiple collections are involved.
However, in the static taint flow analysis, the red lines in 
Code~\ref{lst:el} (Lines $2$-$4$) represent a source, but the data fetched from the source are sent 
to the blue lines (Lines $9$-$10$) which are not a sink. 
In Code~\ref{lst:fe}, the red lines (Lines $9$-$10$) are a sink, but the received data are obtained from the blue
lines (Lines $2$-$4$) which are not a source. Thus, the leak of the sensitive data can be identified
from neither Code~\ref{lst:el} nor Code~\ref{lst:fe}, and the key reason is that the code representing 
the source and sink are modified on purpose to hide the taint flow. AppSpear claims that
it implements an instruction-level tracing mechanism, however, as we will explain below, simply tracing
the instructions does not satisfy the requirement of static analysis tools.

\noindent\textbf{Instruction-level Collection and Tree Model.} 
In light of the shortcoming of method-level collection as described above, the \TheName{} leverages 
instruction-level collection to defend against self-modifying code such as Code~\ref{lst:smc}. 
One simple approach for instruction-level collection is to list all the executed instructions one by one; however, this approach leads to a code scale issue. 
Take the loop as an example, since the instructions
in a loop are executed for multiple times, the simple approach would lead to a large number of repeating
instructions.
Moreover, the branch statements and self-modifying code make it possible that
different executions of a single method lead to different instruction sequences.
However, the format of the DEX file~\cite{dex:gg} allows only one instruction sequence for a single method.

\setlength{\textfloatsep}{0pt}
\setlength{\floatsep}{0pt}
\begin{lstlisting}[caption={Smali representation of the method \texttt{advanced\-Leak} while entering and leaving it.},label={lst:el}, float=tp, language=Ant]
.method public advancedLeak()V
  <@\textcolor{red}{invoke-virtual { p0 }, \textbackslash}@>
  <@\ \ \textcolor{red}{Lcom/test/Main;->getSensitiveData()Ljava/lang/String;}@>
  <@\textcolor{red}{move-result-object v0}@>
  const/4 v1, 0
  :L0
  const/4 v2, 2
  if-ge v1, v2, :L1
  <@\textcolor{blue}{invoke-virtual { p0, v0 }, \textbackslash}@>
  <@\ \ \textcolor{blue}{Lcom/test/Main;->normal(Ljava/lang/String;)V}@>
  invoke-virtual { p0, v1 }, \
  <@\ \ Lcom/ecspride/Main;->bytecodeTamper(I)V@>
  add-int/lit8 v1, v1, 1
  goto :L0
  :L1
  return-void
.end method
\end{lstlisting}

\setlength{\textfloatsep}{5pt}
\begin{lstlisting}[caption={Smali representation of the method \texttt{advance\-dLeak} after the first execution of the
method \texttt{bytecode\-Tamper}.}, float=tp, label={lst:fe}, language=Ant]
.method public advancedLeak()V
  <@\textcolor{blue}{const-string v0, "non-sensitive data"} @>
  <@\textcolor{blue}{nop} @>
  <@\textcolor{blue}{nop} @>
  const/4 v1, 0
  :L0
  const/4 v2, 2
  if-ge v1, v2, :L1
  <@\textcolor{red}{invoke-virtual { p0, v0 }, \textbackslash}@>
  <@\ \ \textcolor{red}{Lcom/test/Main;->sink(Ljava/lang/String;)V}@>
  invoke-virtual { p0, v1 }, \
  <@\ \ Lcom/ecspride/Main;->bytecodeTamper(I)V@>
  add-int/lit8 v1, v1, 1
  goto :L0
  :L1
  return-void
.end method
\end{lstlisting}

To address the code scale issue, \TheName{} eliminates repeating instructions 
by comparing the instructions with same indices.
As mentioned above, the bytecode of a method is organized in a 16-bit unit array and passed to the
interpretation functions (\texttt{ExecuteSwitchImpl} and \texttt{ExecuteGotoImpl} functions).
In these functions, the interpreter uses a variable \texttt{dex\_pc}
to represent the index of the executing instruction in the array. In light of this, we
identify repeating instructions by comparing the executing instructions with the same \texttt{dex\_pc} values. Moreover, the self-modifying
code can also be identified by the comparison. Different instructions
with the same \texttt{dex\_pc} value actually indicate a runtime modification.

\begin{figure}[t]
\centering
\includegraphics[width=3.4in]{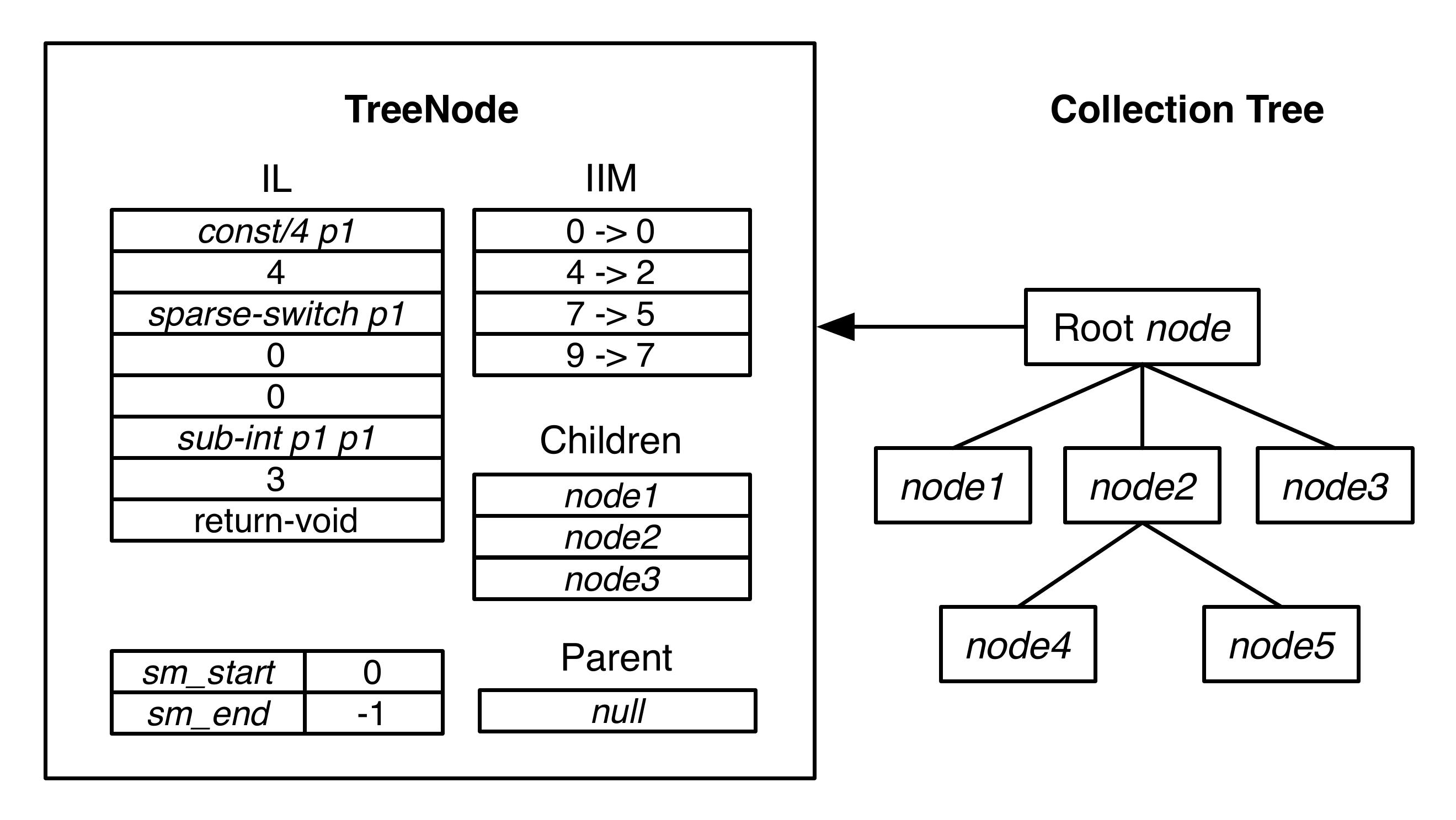}
\caption{Data Structure Storing All Instructions in a Method During a Single Execution. The right
tree structure shows the collection result for a method during a single execution. The left rectangle
describes the data structure of each tree node. For each execution of a method, we generate a collection tree.}
\label{fig:tn}
\end{figure}

Algorithm~\ref{alg:riea} illustrates the comparison-based instruction collection algorithm, and
Figure~\ref{fig:tn} shows the related data structures. We consider the first execution of an instruction as 
a baseline and any different instructions with the same \texttt{dex\_pc} value as a divergence branch. 
Thus, each divergence branch indicates a piece of self-modifying code. Note that self-modifying code
might also exist in the divergence branch (like multiple layers of self-modifying). 
The divergence branches in a method then form a tree structure. The right part of Figure~\ref{fig:tn}
shows an example of the final collecting result. Nodes $1$-$3$ represent three pieces of self-modifying code 
on the root node, and Nodes $4$-$5$ represent two pieces of self-modifying code on Node $2$.
The left rectangle in Figure~\ref{fig:tn} shows the \texttt{TreeNode} structure which represents
a node in the tree structure. The Instruction List (IL) in the structure includes the list of
executed instruction and their metadata. The instructions in IL are recorded by the
order of their first execution and the IL plays the role of baseline
in the node. The \texttt{dex\_pc} value of an instruction
may be different from its index in IL due to branch statements, and we use an
Instruction Index Map (IIM) to maintain the mapping between the instruction's \texttt{dex\_pc}
value and its index in IL for further comparison.
\texttt{sm\_start} and \texttt{sm\_end} indicate
the starting and ending \texttt{dex\_pc} value of the divergence branch, while \texttt{parent} and \texttt{children}
represent the parent and all children of the node, respectively. 
With the tree structure, \TheName{} records all executed instructions in a single execution of a method 
and maintains the code size similar to the original instruction array. 

\setlength{\textfloatsep}{10pt}
\begin{algorithm}
\caption{Bytecode Collection Algorithm}
\label{alg:riea}
\begin{algorithmic}[1]
\Procedure{BytecodeCollection}{}
\State create node $root$
\State $current$ = $root$
\For{each executing instruction $ins$}
\State let index of $ins$ be $dex\_pc$
\If{$dex\_pc$ exists in $current.IIM$}
\State $pos\_in\_IL$ = $current.IIM.get(dex\_pc)$
\State $old\_ins$ = $current.IL.get(pos\_in\_IL)$
\If{!SameIns($ins$, $old\_ins$)}
\State create a child node $child$
\State $child.parent$ = $current$
\State $child.start\_pos$ = $dex\_pc$
\State $current$ = $child$
\Else
\State $continue$
\EndIf
\ElsIf{$current$ has a parent}
\State $parent$ = $current.parent$
\If{$dex\_pc$ exists in $parent.IIM$}
\State $pos\_in\_IL$ = $parent.IIM.get(dex\_pc)$
\State $old\_ins$ = $parent.IL.get(pos\_in\_IL)$
\If{SameIns($ins$, $old\_ins$)}
\State $current.end\_pos$ = $dex\_pc$
\State $current$ = $parent$
\State $continue$
\EndIf
\EndIf
\EndIf
\State $pos\_in\_IL$ = $current.IL.size()$
\State $current.IL.add(ins)$
\State $current.IIM.push(pair(dex\_pc, pos\_in\_IL))$
\EndFor
\EndProcedure
\end{algorithmic}
\end{algorithm}

In Algorithm~\ref{alg:riea}, we only update one node during
the execution of a single instruction, and this node is considered as the
current node. \TheName{} creates an empty root node as the current node while entering a method.
Once an instruction is executed, we check IIM 
of the current node to find whether the \texttt{dex\_pc} value of this instruction has been recorded. 
If it does not exist in IIM, \TheName{} pushes the instruction into IL and updates IIM. If the
\texttt{dex\_pc} value already exists in IIM, we add a check procedure to find whether the instruction 
is the same as the one we recorded before. A positive result means that the same instruction in the same 
position is executed again, and \TheName{} does not record it. In contrast, the negative 
result indicates that modification has occurred to this instruction since its last execution. 
Then, we create a child node of the current node to represent the divergence branch, 
and the new node becomes the current node. After that, 
\TheName{} treats the instruction as a new instruction and pushes it into IL of the current node. 
In a divergence branch, another check procedure is added to each
instruction, and this check procedure aims to identify whether the current divergence
branch converges to its parent. If the same instruction with the same \texttt{dex\_pc} value has been found in
the parent's IL, we consider that the divergence branch converges back to its parent (e.g., current 
layer of self-modifying code ends) and make the parent node to be the new current node.

Listing~\ref{lst:drsmc} shows a high-level semantic view of the collection result of the method \texttt{advancedLeak} in 
Code~\ref{lst:smc}. When Line $13$ in Code~\ref{lst:smc} is executed for the first time, an invocation of the method \texttt{normal} 
is recorded. Then, in the second run, an invocation of the method \texttt{sink} is detected. However, 
by comparing with the recorded instructions, \TheName{} finds that it is a divergence point. A 
child node is forked and the instruction is pushed into the IL of the child node. Furthermore, 
a convergence point is found when Line $14$ is executing. Thus, the collection tree contains a root node 
and a child node, and the child node contains only one instruction. With the tree, the executed instructions
and the control flows in the method are well maintained. Note that the modification to the Line $11$ is ignored
since the modified instructions are never executed.

\setcounter{lstlisting}{0}
\renewcommand{\lstlistingname}{Listing}
\setlength{\textfloatsep}{3pt}
\begin{lstlisting}[label={lst:drsmc}, float=tp, caption={High-level Semantic View of the Collection Result of 
the Method \texttt{advancedLeak} in Code~\ref{lst:smc}.}]
Root Node:
  String a = getSensitiveData();
  for (int i = 0; i < 2; ++i) {
    normal(a);
    bytecodeTamper(i); 
  }

Child Node: (Line 13 in Code 1)
  sink(a);
\end{lstlisting}

\renewcommand{\lstlistingname}{Code}
\setcounter{lstlisting}{3}

For the issue of multiple instruction sequences for a single method, we generate multiple
collection trees for multiple executions of the method and keep only the unique trees.
The trees are further combined together with the approach detailed in Section~\ref{ssec:bcr}. 

\subsection{Bytecode Reassembling}
\label{ssec:bcr}

The offline reassembling-phase merges the collected trees into a DEX file while
holding all the executed instructions and control flows. There are two steps in this phase: 
1) converting each tree into an instruction array.
2) merging instruction arrays into the DEX file.


\noindent\textbf{Converting a Tree into an Instruction Array.} Each node in the collection tree generated from the collection phase
contains an independent Instruction List (IL), and the goal of this phase is to
combine the ILs in the nodes together without losing any control flows or instructions. 
To simplify the combination process, we traverse the nodes with the bottom-up fashion 
since the leaf nodes contain no child node.

To merge a single leaf to its parent,
\TheName{} inserts an additional branch instruction in the divergence point 
(indicated by \texttt{sm\_start}, self-modifying start, as defined in the above subsection~\ref{ssec:bcd}), 
with one branch of the instruction pointing to the leaf. 
To make both conditional branches reachable, the  
conditional expression of the added
branch instruction is calculated based on a static field of an instrument class
with random values.
Note that the random value produces indeterminacy problem on the additional
branch instruction, and we consider it acceptable since the static analysis tool
will take both branches of the instruction as reachable.

Once the leaf nodes are recursively merged into their parents, the root node
becomes a complete set of the collected instructions including different
control flows triggered during the execution.

Code~\ref{lst:rrsmc} demonstrates the reassembled result of 
Listing~\ref{lst:drsmc}. The static field \texttt{com\_test\_Main\_advancedLeak\_0}
in our instrument class \texttt{Mo\-dification} indicates the divergence point in
Line $13$ of Code~\ref{lst:smc}. When this result is fed to static analysis tools, they
treat both \texttt{normal} and \texttt{sink} as reachable and detect
the taint flow from sensitive data to text message in Code~\ref{lst:smc}.

\begin{lstlisting}[label={lst:rrsmc}, float=tp, caption={Reassembled Result of the 
Method \texttt{advancedLeak} in Code~\ref{lst:smc}.}]
String a = getSensitiveData();
for (int i = 0; i < 2; ++i) {
  if (Modification.com_test_Main_advancedLeak_0) {
    normal(a);
  } else {
    sink(a)
  }
  bytecodeTamper(i);
}
\end{lstlisting}

\noindent\textbf{Merging Instructions Arrays.} For each executed method, the previous phase outputs unique instruction arrays which
indicate different executions of the method. Similar to the approach discussed above,
we create a method variant for each instruction array and use additional branch
instructions to cover different method variants.

\subsection{Data Collection and DEX Reassembling}

As mentioned in Section~\ref{ssec:abdc}, besides bytecode instructions, \TheName{} uses JIT 
collection to collect the metadata of DEX file. The collected data is written into
collection files and further used to reassemble a new DEX file offline.

In Code~\ref{lst:smc}, before any method or field in \texttt{Main} 
is accessed, the class \texttt{Lcom/example/Main;} is loaded and initialized. During the process, we
firstly store string \texttt{Lcom/example/Main;} into a \texttt{st\-ring} structure and
record the index of this \texttt{string} structure. Then with the index,
a \texttt{type} structure is constructed and stored. Finally, a corresponding \texttt{class} structure
related to the \texttt{type} is extracted. The collection occurs again when the class is 
initialized. The initialization procedure links the methods and fields to the class, and initializes the static
fields. In Code~\ref{lst:smc}, methods \texttt{onCreate}, \texttt{advancedLeak}, \texttt{normal}, and 
\texttt{sink} are linked to the class. While the static field \texttt{PHONE} is 
initialized, \TheName{} stores its name \texttt{PHONE}, type \texttt{Ljava/lang/String;}
and initial value \texttt{800-123-456}. Lastly, a \texttt{field} structure is created and 
recorded. The \texttt{method} structures and the bytecode inside them are collected before
and during the execution of the methods, respectively.

After the collection process, all collection files including bytecode are combined offline according to the
format of the DEX file. Finally, we leverage the Android Asset Packaging Tool
integrated with Android SDK to replace the DEX file in the original APK file with the reassembled one. 
To verify the soundness of our extracting and reassembling algorithm, we perform extensive tests
against real-world applications, and the evaluation results in Section~\ref{ssec:packer},
Section~\ref{ssec:twdb}, and Section~\ref{ssec:cceval} show that the 
reassembled DEX file retains the semantics of the real-world application and can be 
correctly processed by the state-of-the-art static analysis tools.

\subsection{Handling Reflection}
\label{ssec:er}

Currently, reflection is a serious obstacle for static analysis tools, 
and even the state-of-the-art static analysis tools~\cite{flowdroid:pldi14,horndroid:eurosp16,droidsafe:ndss15,harvester:ndss16} 
cannot provide a precise result when reflection is involved in an application. 
FlowDroid~\cite{flowdroid:pldi14}, DroidSafe~\cite{droidsafe:ndss15}, and 
HornDroid~\cite{horndroid:eurosp16} can solve the reflection only
when the parameters are constant strings. However, the name string can be encrypted in some cases~\cite{harvester:ndss16}, and advanced malware can use reflection without involving any string parameter~\cite{droidbench:ecssse}.

The TamiFlex~\cite{tamiflex:icse11} system on traditional Java platform
uses load-time instrumentation to log reflective method calls and transform
them to direct calls at offline. However, the
required load-time instrumentation class \texttt{java.lang.instrument} is
not supported in Android~\cite{flowdroid:pldi14}. Meanwhile, since
the target of the reflective method calls is parsed in ART at runtime, \TheName{}
actually knows the target of each reflection. Thus, we apply the similar
idea in ART by replacing the reflection calls with direct calls in the collecting stage.

\subsection{Force Execution}
\label{ssec:cc}

\begin{figure}
\centering
\includegraphics[width=3.2in]{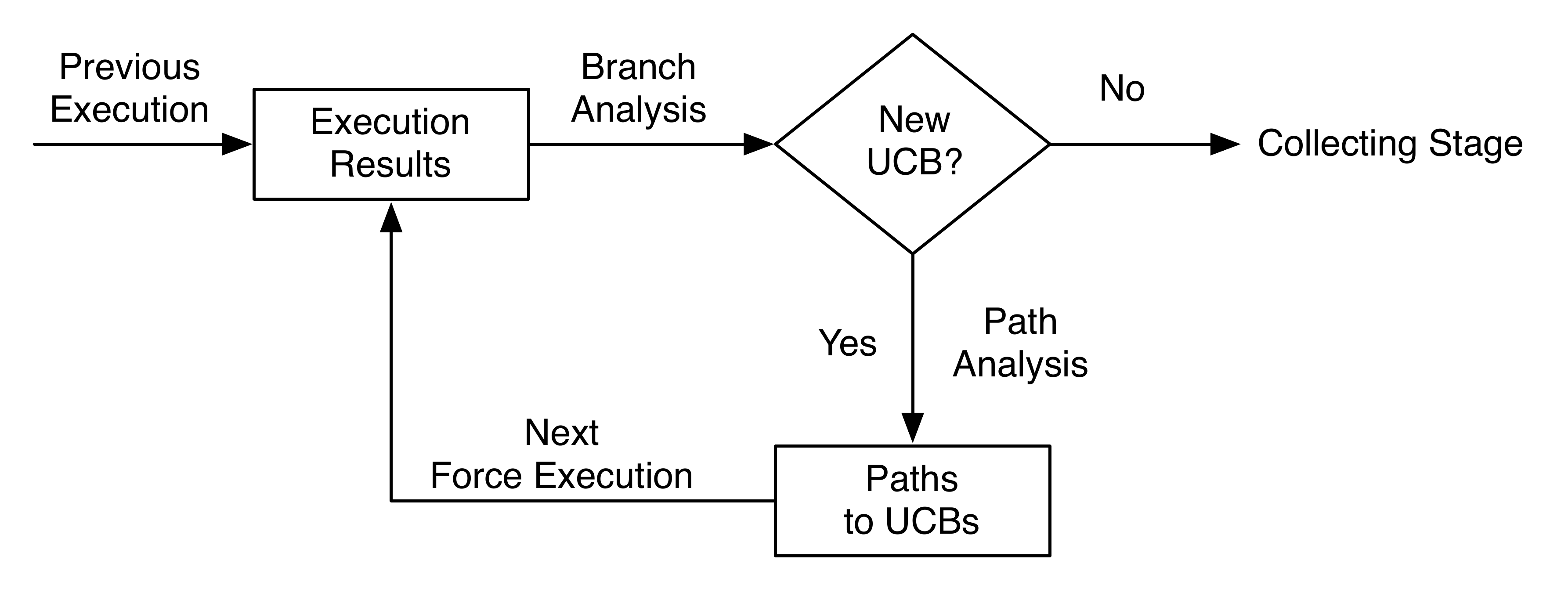}
\caption{Iterative Force Execution.}
\label{fig:forceexecution}
\end{figure}

As a supplement of the code coverage improvement module, we implement a prototype of
force execution which executes the target application in an iterative fashion. 
Note that our force execution starts from the execution result of the previous execution,
and the previous execution could be any kind of execution like fuzzing, symbolic
execution, another force execution, or simply open the application and close.
Figure~\ref{fig:forceexecution} shows the workflow of the iterative force execution.
In each iteration, we first use branch analysis to identify
the Uncovered Conditional Branch (UCB) from the result of the previous execution.
Next, we calculate
the control flow path to each UCB. A path to an UCB consists of branch instructions
and the offsets of the conditional branches leading to the UCB. We save each path
into a file and use these files as the input of the next iteration together with
the original application. Finally, in the interpretation functions, the outcome
of the corresponding conditional expression is automatically manipulated at
runtime following the path files. With this approach, \TheName{} ensures that
the runtime control flow goes along the path to the UCB. If no more new UCB are generated after the 
iteration, we terminate the execution
and continue the collecting stage. Otherwise, the next iteration is scheduled. 

Since the idea of force execution breaks the normal control flow of the original application, the application
may crash due to the control flow falls to an infeasible path~\cite{jforce:www17,xforce:usenix14}. To avoid 
crash triggered by force execution, we monitor the unhandled exception in the interpreter and tolerate it by directly 
clear the exception. 
This strategy helps us to avoid terminations due to infeasible paths while does not affect our runtime 
bytecode and data collection. 
\section{Evaluation}
\label{sec:evaluation}

In this section, we evaluate \TheName{} with DroidBench~\cite{droidbench:ecssse}
and real-world applications downloaded from Google Play and other application markets.
In particular, we aim to answer five research questions:

\noindent\textbf{RQ1.} Can we correctly reconstruct the behavior of apps? \\
\noindent\textbf{RQ3.} How is \TheName{} compared with other tools? \\
\noindent\textbf{RQ4.} Can \TheName{} work with real-world packed apps? \\
\noindent\textbf{RQ5.} What is the coverage of our force execution prototype? \\
\noindent\textbf{RQ6.} What is the runtime performance overhead?


\subsection{RQ1: Test with Open-source Apps and Public Packers}
\label{ssec:packer}

To verify the correctness of the reassembled result, we pick up four open source
applications (i.e., \texttt{HTMLViewer}, \texttt{Calculator}, \texttt{Calendar}, 
and \texttt{Contacts}) from AOSP~\cite{asop:gg} and use \TheName{} 
to reveal them. By manually comparing the instructions and control flows
in each method, we ensure that the instructions and control flows in the
source code are completely included in the reassembled result. In regard to 
\texttt{Calendar} and \texttt{Contacts}, we use Soot framework~\cite{soot:cetus11} 
to build a complete call graph since the numbers of instructions (78,598 and 103,602 instructions,
respectively) are too large for a manual analysis. By examining the call graph, we confirm that
the control flows in these two applications are properly maintained in the reassembled DEX.

Next, to check the functionality against packers, we use different public packing platforms
to pack these applications and then use \TheName{} to reveal them again. 
Table~\ref{table:trdp} shows the result of the experiments. 
For the packers including 360~\cite{360:protector}, Alibaba~\cite{ali:protector}, 
Tencent~\cite{tencent:protector}, Baidu~\cite{baidu:protector}, and 
Bangcle~\cite{bangcle:protector}, \TheName{} succeeds 
in both collection and reassembling stages. By using the same approach described above,
we ensure that \TheName{} correctly rebuilds the behavior of each application.
Note that NetQin packer~\cite{netqin:protector} mentioned in AppSpear~\cite{appspear:raid15}
is no longer available. The APKProtect~\cite{apkprotect:protector} is unresponsive to 
the packing requests, and there are no logs of the occurred errors. The packing service
provided by Ijiami~\cite{ijiami:protector} requires manual audits by their agents, and 
they reject our applications for the reason that the applications are not actually 
developed by us. 

\setlength{\abovecaptionskip}{0pt}
\setlength{\textfloatsep}{5pt}
\begin{table}
\centering
\caption{Test Result of Different Packers.}
\scriptsize
\begin{tabular}{lcccc} \\
\toprule
  \multicolumn{1}{ l }{Applications} 
  & \multicolumn{1}{ c }{HTMLViewer} 
  & \multicolumn{1}{ c }{Calculator} 
  & \multicolumn{1}{ c }{Calendar} 
  & \multicolumn{1}{ c }{Contacts} \\ 
\midrule
  \# of Instructions & 217 & 2,507 & 78,598 & 103,602 \\
\midrule
  \multicolumn{1}{ l }{360~\cite{360:protector}} 
  & \multicolumn{1}{ c }{\checkmark} 
  & \multicolumn{1}{ c }{\checkmark} 
  & \multicolumn{1}{ c }{\checkmark} 
  & \multicolumn{1}{ c }{\checkmark} \\
  \multicolumn{1}{ l }{Alibaba~\cite{ali:protector}} 
  & \multicolumn{1}{ c }{\checkmark} 
  & \multicolumn{1}{ c }{\checkmark} 
  & \multicolumn{1}{ c }{\checkmark} 
  & \multicolumn{1}{ c }{\checkmark} \\
  \multicolumn{1}{ l }{Tencent~\cite{tencent:protector}} 
  & \multicolumn{1}{ c }{\checkmark} 
  & \multicolumn{1}{ c }{\checkmark} 
  & \multicolumn{1}{ c }{\checkmark} 
  & \multicolumn{1}{ c }{\checkmark} \\
  \multicolumn{1}{ l }{Baidu~\cite{baidu:protector}} 
  & \multicolumn{1}{ c }{\checkmark} 
  & \multicolumn{1}{ c }{\checkmark} 
  & \multicolumn{1}{ c }{\checkmark} 
  & \multicolumn{1}{ c }{\checkmark} \\ 
  \multicolumn{1}{ l }{Bangcle~\cite{bangcle:protector}} 
  & \multicolumn{1}{ c }{\checkmark} 
  & \multicolumn{1}{ c }{\checkmark} 
  & \multicolumn{1}{ c }{\checkmark} 
  & \multicolumn{1}{ c }{\checkmark} \\ 
  \multicolumn{1}{ l }{NetQin~\cite{netqin:protector}} 
  & \multicolumn{4}{ l }{The service is offline now} \\  
  \multicolumn{1}{ l }{APKProtect~\cite{apkprotect:protector}} 
  & \multicolumn{4}{ l }{Unresponsive to packing requests} \\ 
  \multicolumn{1}{ l }{Ijiami~\cite{ijiami:protector}} 
  & \multicolumn{4}{ l }{Samples are rejected by human agents} \\ 
\bottomrule
\end{tabular}
\label{table:trdp}  
\end{table}

\subsection{RQ2: Test with Existing Tools}
\label{ssec:twdb}

\subsubsection{Static Analysis Tools}

DroidBench~\cite{droidbench:ecssse} is a set of open-source samples that
leak sensitive data in various ways. It is considered as a benchmark for Android
application analysis and widely used among recent analysis tools~\cite{flowdroid:pldi14,horndroid:eurosp16,droidsafe:ndss15,iccta:icse15,amandroid:ccs14}. The 
latest release of DroidBench contains 119 applications, including both leaky and benign
samples. The leaky samples leak a variety of sensitive data fetched from sources (API calls
that fetch sensitive information) to sinks (API calls that may leak information), and the
benign samples contain no such information flows. As a supplement, we contribute another 
15 samples
involving usage of advanced reflection (5 samples), 
dynamic loading (3 samples), self-modifying (4 samples), and unreachable taint flows (3 samples). 
Current static analysis tools~\cite{flowdroid:pldi14,horndroid:eurosp16,droidsafe:ndss15} cannot 
precisely analyze these newly added samples. Besides this benchmark, we choose three 
representative static analysis tools (FlowDroid~\cite{flowdroid:pldi14}, 
DroidSafe~\cite{droidsafe:ndss15}, and 
HornDroid~\cite{horndroid:eurosp16}) to conduct the experiments.

Our experiment involves $134$ samples ($119$ samples in the newest release plus $15$ 
samples we contributed) in DroidBench. Since the lines of code in DroidBench samples are 
small, we simply choose the state-of-the-art fuzzing tool Sapienz~\cite{sapienz:issta16} to 
generate the inputs for the execution. We first use the static
analysis tools to analysis the original samples and the samples processed by \TheName{},
and the result is shown in Table~\ref{table:arsat}. The table shows that \TheName{} increases more than $8$
true positives by resolving advanced reflections, extracting self-modifying code and dynamic 
loading code. Moreover, The JIT collection ensures
that the extracted data reflects the performed behavior of the target application. Thus,
at least $5$ false positives introduced by dead code blocks are removed.
Next, without losing generality, we use one of the most popular packers tested in 
Section~\ref{ssec:packer}, 360 packer, to pack the original samples and process the
packed samples with \TheName{}, DexHunter~\cite{dexhunter:esorics15}, and AppSpear~\cite{appspear:raid15}, 
respectively. The analysis
result of the processed samples is shown in Table~\ref{table:arps}. 
Note that DexHunter and AppSpear lead to the same result since they can extract
the original DEX files and the result is same as analyzing the original DEX. Compared to \TheName{},
they fail to deal with self-modifying code and reflection.
As shown in the table, \TheName{}
provides more than $5$ true positives and reduces more than $5$ false positives than
DexHunter and AppSpear.
We note that \TheName{} fails to cover taint flow in only one application 
among all samples. In this sample, sensitive data only leaks in the tablet, and it cannot be 
detected as we execute it in a mobile phone.

\setlength{\floatsep}{10pt}
\begin{table}
\scriptsize
\centering
\caption{Analysis Result of Static Analysis Tools. The columns in "Original" represent the 
analysis result of the original samples, and the columns in "\TheName{}"
represent that of the samples reassembled by \TheName{}. The column "TP"
and "FP" indicate the number of true positives and false positives of the 
analysis result, respectively.}
\begin{tabular}{lcccccc} \\
\toprule
  & 
  \multirow{2}{*}[-0.3em]{\makecell{\# of \\ Samples}} &
  \multirow{2}{*}[-0.3em]{\makecell{\# of \\ Malware}} &
  \multicolumn{2}{c}{Original} & 
  \multicolumn{2}{c}{\TheName{}} \\
\cmidrule(r){4-5}
\cmidrule(r){6-7}
  & & & 
  TP & FP & 
  TP & FP \\
\midrule
  FlowDroid~\cite{flowdroid:pldi14} & 134 & 111 & 81 & 10 &   95  & 4 \\
  DroidSafe~\cite{droidsafe:ndss15} & 134 & 111 & 95 & 12 &  105 & 7 \\
  HornDroid~\cite{horndroid:eurosp16} & 134 & 111 & 98 &   9 & 106  & 4 \\
\bottomrule
\end{tabular}
\label{table:arsat}
\end{table}

\setlength{\floatsep}{10pt}
\begin{table}
\scriptsize
\centering
\caption{Analysis Result of Packed Samples. The columns in "DH",
"AS", and "\TheName{}" represent the analysis result of the
samples processed by DexHunter~\cite{dexhunter:esorics15},
AppSpear~\cite{appspear:raid15}, and \TheName{}, respectively. The column "TP"
and "FP" indicate the number of true positives and false positives of the 
analysis result, respectively.}
\begin{tabular}{lcccccc} \\
\toprule
  & 
  \multirow{2}{*}[-0.3em]{\makecell{\# of \\ Samples}} &
  \multirow{2}{*}[-0.3em]{\makecell{\# of \\ Malware}} &
  \multicolumn{2}{c}{DH~\cite{dexhunter:esorics15} / AS~\cite{appspear:raid15}} &
  \multicolumn{2}{c}{\TheName{}} \\
\cmidrule(r){4-5}
\cmidrule(r){6-7}
  & & & 
  \multicolumn{1}{c}{TP} & 
  \multicolumn{1}{c}{FP}  &
  TP & FP \\
\midrule
  FlowDroid~\cite{flowdroid:pldi14} & 134 & 111 &  84 & 10 & 95 & 4\\
  DroidSafe~\cite{droidsafe:ndss15} & 134 & 111 &  98 & 12 & 105 & 7 \\
  HornDroid~\cite{horndroid:eurosp16} & 134 & 111 & 101 &  9 & 106 & 4\\
\bottomrule
\end{tabular}
\label{table:arps}
\end{table}

\vspace{-10pt}
\begin{align}
\scriptsize
\centering
\label{align:stats}
\begin{split}
  Sensitivity = \frac{tp}{tp + fn},\ Specificity = \frac{tn}{tn + fp}, \\
  \textit{F-Measure} = 2 \times \frac{Sensitivity \times Specificity}{Sensitivity + Specificity}
\end{split}
\end{align}  

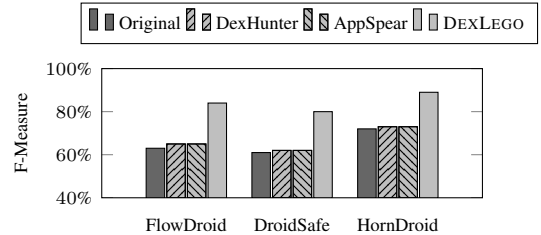
\begin{figure}[t]
\centering
\begin{tikzpicture}
      \begin{axis}[
      width  = 8cm,
      height = 3.3cm,
      font=\scriptsize,
      major x tick style = transparent,
      ybar=2*\pgflinewidth,
      x = 40pt,
      bar width=7pt,
      symbolic x coords={FlowDroid, DroidSafe, HornDroid},
      xtick = data,
      yticklabel={\pgfmathparse{\tick*100}\pgfmathprintnumber{\pgfmathresult}\%},
      scaled y ticks = false,
      enlarge x limits=0.40,
      ymin=0.4,
      ylabel={F-Measure},
      ymax=1,
      legend cell align=left,
      legend style={legend columns=4, at={([yshift=25pt]0,1)}, xshift=-8pt ,anchor=north west,
cells={anchor=west}},
  ]
      \addplot[style={fill=darkgray!80}]
          coordinates {
          (FlowDroid, 0.63)
          (DroidSafe,0.61)
          (HornDroid,0.72)};
      \addplot[style={fill=lightgray}, postaction={pattern=north east lines}]
           coordinates {
           (FlowDroid,0.65)
           (DroidSafe,0.62) 
           (HornDroid,0.73)};
      \addplot[style={fill=lightgray}, postaction={pattern=north west lines}]
           coordinates {
           (FlowDroid,0.65)
           (DroidSafe,0.62) 
           (HornDroid,0.73)};
	  \addplot[style={fill=lightgray}]
           coordinates {
           (FlowDroid,0.84)
           (DroidSafe,0.80) 
           (HornDroid,0.89)};

      \legend{\scriptsize{Original}, \scriptsize{DexHunter},
         			 \scriptsize{AppSpear}, \scriptsize{\TheName{}}}
  \end{axis}
  \end{tikzpicture}
\caption{F-measures of Static Analysis Tools.}
\label{fig:fart}
\end{figure}

The F-Measure~\cite{horndroid:eurosp16} is a standard measure of the performance of a classification, and it is
calculated by Formula~\eqref{align:stats}. Figure~\ref{fig:fart} illustrates the changes of 
F-Measures after involving DexHunter, AppSpear,
and \TheName{}. Once \TheName{} is involved, the F-Measure of FlowDroid increases from 
$63\%$ to $84\%$ on DroidBench, and that of DroidSafe increases from $61\%$ to $80\%$. In 
regard to the most recent static analysis tool, HornDroid, the F-Measure increases 
from $72\%$ to $89\%$. The percentages 
of incremental values are $33.3\%$, $31.1\%$, and $23.6\%$, respectively. In the meantime,
the improvement introduced by DexHunter and AppSpear is less than $3\%$.

\setlength{\textfloatsep}{5pt}

\subsubsection{Dynamic Analysis Tools}

As mentioned in Section~\ref{sec:intro}, dynamic analysis tools can be circumvented
through implicit taint flows, and a recent work~\cite{harvester:ndss16} shows that 
a representitive dynamic analysis tool, TaintDroid~\cite{taintdroid:osdi10},
misses leakage on some samples of DroidBench. We pick these samples and analyze
them with both TaintDroid and another recent dynamic analysis tool 
TaintART~\cite{taintart:ccs16}. Next, we use \TheName{} to analyze it again. The
reassembled result is fed to HornDroid, the most recent static analysis tool,
for comparison.

Table~\ref{table:ardat} shows the taint flow analysis results of TaintDorid, TaintART, 
and combing \TheName{} and HornDroid. 
As shown in Table~\ref{table:ardat}, the static analysis result of reassembled APK
file by \TheName{} detects the taint flows and is more precise than dynamic analysis 
tools. In \texttt{Button1} and \texttt{Button3}, the sensitive data
are leaked via callback methods, and we solve it properly while
the dynamic analysis tools miss it. As TaintDroid executes applications on emulator, the sample
\texttt{EmulatorDetection1} evades the analysis. Both TaintDroid and TaintART cannot
detect the implicit taint flows in \texttt{ImplicitFlow1}, and using HornDroid with \TheName{} 
provides a precise
analysis result. One of the taint flows in \texttt{PrivateDataLeak3} leaks the sensitive data
through writing/reading an external file, and all tested tools fail to detect this flow since
they do not take this case into account. Note that these missed taint flows are not
caused by code coverage issue, but due to the weakness of dynamic analysis tools
on implicit taint flows.

\setlength\tabcolsep{3pt}
\begin{table}
\centering
\caption{Analysis Result of Dynamic Analysis Tools and \TheName{}. 
The columns "TD" and 
"TA" represent the taint flows detected by TaintDroid~\cite{taintdroid:osdi10}
and TaintART~\cite{taintart:ccs16}, respectively. The last column shows the detected taint flows
by feeding the revealed result of \TheName{} to HornDroid~\cite{horndroid:eurosp16}.}
\scriptsize
\begin{tabular}{lcccc} \\
\toprule
  \multicolumn{1}{ c }{\multirow{2}{*}{Samples}}
  & \multicolumn{1}{ c }{\multirow{2}{*}{Leak \#}}
  & \multicolumn{3}{ c }{\# of Leak Detected} \\ 
\cmidrule(r){3-5}
  &
  & \multicolumn{1}{ c }{TD~\cite{taintdroid:osdi10}}
  & \multicolumn{1}{ c }{TA~\cite{taintart:ccs16}}
  & \multicolumn{1}{ c }{\TheName{} + HD~\cite{horndroid:eurosp16}} \\
\midrule
  Button1 & 1 & 0 & 0 & 1 \\
  Button3 & 2 & 0 & 0 & 2 \\
  EmulatorDetection1 & 1 & 0 & 1 & 1 \\
  ImplicitFlow1 & 2 & 0 & 0 & 2 \\
  PrivateDataLeak3 & 2 & 1 & 1 & 1 \\
\bottomrule
\end{tabular}
\label{table:ardat} 
\end{table}

Note that \TheName{} is not a dynamic analysis tool. We believe we should not directly compare \TheName{} with dynamic analysis tools, and the dynamic analysis tools have their advantages. However, the experiment conducted in this subsection is to show that \TheName{} can help static analysis tools make up some deficiencies of dynamic analysis tools.

\subsection{RQ3: Test with Real-world Packed Applications}
\label{ssec:trwa}

A previous work~\cite{playdrone:sigmetrics14} has downloaded more than one million 
applications from Google Play by a crawler in 2014, and we select the packed applications
from this set. 
Since the DEX file in an application packed by the public packing platforms contains only the classes
needed to unpack the original DEX file, the number of the classes in the DEX file is less compared to
normal applications. In light of this, we perform a coarse-grain analysis to screen the applications 
which contains less than $50$ classes from the top rated $10,000$ applications. 
Next, we select the first $9$ applications from the screened result by manually checking
and reverse engineering. Without loss of generality, we download the latest version
of these applications from three different popular application markets: 1) Google 
Play~\cite{googleplay:gg} (denoted as set \texttt{A}), 
2) 360 Application Market~\cite{360market:360}
(denoted as set \texttt{B}), and 3) Wandoujia Application Market~\cite{wandoujia:wandoujia}
(denoted as set \texttt{C}).

For these real-world packed applications, we use FlowDroid to provide a quick scan on the original
applications, and then execute them with \TheName{} for $5$ minutes. Next, the reassembled
APK file is analyzed again by FlowDroid. Table~\ref{table:arpra} shows the result of our
experiment. Although no taint flow can be detected from the original samples, FlowDroid
detects several taint flows from these revealed applications. From the analysis result, we find that 
all of these applications send device ID (IMEI number) to remote servers. Moreover, three
of them leak location information and two of them leak SSID. This result also shows that
\TheName{} successfully reveals the latest packed real-world applications.

\setlength\tabcolsep{2pt}
\begin{table}
\centering
\caption{Analysis Result of Packed Real-world Applications. 
The column "Sample Set" is
defined in Section~\ref{ssec:trwa}, which indicates the source of the application. 
The column "\# of Installs" shows the installation number provided by the application
markets. The column "Original" represents the number of detected taint flows in the original 
application while the column "Revealed" is the number of detected taint flows in the revealed APK
file.}
\scriptsize
\begin{tabular}{llclcc} \\
\toprule
  \multicolumn{1}{ c }{Package Name}
  & \multicolumn{1}{ c }{Version}
  & \multicolumn{1}{ c }{Sample Set}
  & \multicolumn{1}{ c }{\# of Installs} 
  & \multicolumn{1}{ c }{Original}
  & \multicolumn{1}{ c }{Revealed} \\
\midrule
  com.lenovo.anyshare & 
  3.6.68 & 
  A & 
 100 million &
  0 &
  4 \\
  com.moji.mjweather & 
  6.0102.02 & 
  A & 
  1 million &
  0 &
  5 \\
  com.rongcai.show & 
  3.4.9 & 
  A & 
  100 thousand&
  0 &
  3 \\  
\midrule
  com.wawoo.snipershootwar & 
  2.6 & 
  B & 
  10 million &
  0 &
 4 \\
  com.wawoo.gunshootwar & 
  2.6 & 
  B & 
  10 million &
  0 &
  5 \\
  com.alex.lookwifipassword & 
  2.9.6 & 
  B & 
  100 thousand &
  0 &
  2 \\
\midrule
  com.gome.eshopnew & 
  4.3.5 & 
  C & 
  15.63 million &
  0 &
  3 \\
  com.szzc.ucar.pilot & 
  3.4.0 & 
  C & 
  3.59 million &
  0 &
  5 \\
  com.pingan.pabank.activity & 
  2.6.9 & 
  C & 
  7.9 million &
  0 &
  14 \\
\bottomrule
\end{tabular}
\label{table:arpra}
\end{table}

\subsection{RQ4: Code Coverage}
\label{ssec:cceval}

To evaluate the code coverage of our force execution engine, we pick up 
five open source applications from the random page~\cite{fdroidrandom:fdroid}
of F-Droid~\cite{fdroid:fdroid} project. For each application, we first execute
it with Sapienz~\cite{sapienz:issta16} and use Java Code Coverage Library (JaCoCo)~\cite{jacoco:jacoco}
for Android Studio to calculate the coverage. Next, based on the result of
Sapienz, we execute it again using the force execution engine as the code
coverage improvement module.


Table~\ref{table:sfdroid} shows the details of the samples including package name, version number, 
the number of instructions, and the total size of the dump files after fuzzing by Sapienz. Note
that the size of the dump files is not only related to the number of the instructions in the application,
but also related to the size of other data structures in the DEX file (e.g., number of classes,
number of methods, size of strings, and so on.) and the code coverage of the fuzzing.
Table~\ref{table:ccfdroid} shows the average coverage of these samples with
different granularities. The results show that the force execution significantly improves the coverage
and achieves an average instruction coverage of 82\%.
By manually check the source code, we group the cause of missed instructions into three 
main categories: 1) Dead code blocks. As an example, the
\texttt{CmdTemplate} class is never
involved in the application \texttt{be.ppareit.swiftp}, thus the entire instructions in this class are 
not included while calculating coverage.
2) Native crashes. Although \TheName{} clears the unhandled exceptions in the interpreter,
the abnormal control flows may lead the native code to crash. This may be mitigated by
the on demand runtime memory allocation mechanism applied in~\cite{xforce:usenix14}.
3) Instructions in exception handlers. During force execution, the expected exceptions
in the \texttt{try-catch} blocks may not be thrown due to abnormal control flow, and
it may be solved by treating these blocks as branch instructions in
the branch analysis. We leave it as a future work. 

\setlength\tabcolsep{3pt}
\setlength{\floatsep}{10pt}
\begin{table}
\centering
\caption{Samples from F-Droid~\cite{fdroid:fdroid}.}
\scriptsize
\begin{tabular}{llcc} \\
\toprule
  Package Name & Version & \# of Instructions & Dump File Size\\
\midrule
  be.ppareit.swiftp & 2.14.2 & 8,812 & 47.26 KB \\
  fr.gaulupeau.apps.InThePoche & 2.0.0b1 & 29,231 & 771.81 KB \\
  org.gnucash.android & 2.1.7 & 56,565 & 2.40 MB \\
  org.liberty.android.fantastischmemopro & 10.9.993 & 57,575 & 1.55 MB \\
  com.fastaccess.github & 2.1.0 & 93,913 & 3.18 MB \\
\bottomrule
\end{tabular}
\label{table:sfdroid}
\end{table}

\begin{table}
\centering
\caption{Code Coverage with F-Droid Applications.}
\scriptsize
\begin{tabular}{lccccc} \\
\toprule
 & Class & Method & Line & Branch & Instruction\\
\midrule
  Sapienz~\cite{sapienz:issta16} & 44\% & 37\% & 32\% & 20\% & 32\% \\
  Sapienz + \TheName{} & 87\% & 88\% & 82\% & 78\% & 82\% \\
\bottomrule
\end{tabular}
\label{table:ccfdroid}
\end{table}

\subsection{RQ5: Performance}
\label{ssec:pe}

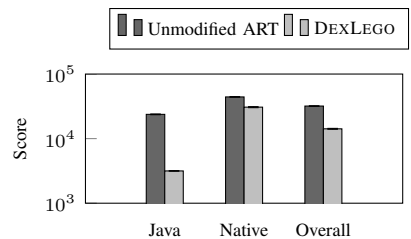
\begin{figure}
\centering
\begin{tikzpicture}
      \begin{axis}[
      width  = 5cm,
      height = 3.3cm,
      font=\scriptsize,
      ybar=0,
      x = 30pt,
      bar width=7pt,
      minor y tick style = transparent,
      major x tick style = transparent,
      xtick = {1, 2, 3},
      xticklabels = {Java, Native, Overall},
      ytick = {1000, 10000, 100000},
      ytick pos = left,
      ymin = 1000,
      ymax = 100000,
      ymode = log,
      enlarge x limits=0.50,
      ylabel={Score},
      ylabel near ticks,
      legend cell align=left,
      legend style={legend columns=2, at={([yshift=25pt]0,1)}, xshift=9pt ,anchor=north west,
cells={anchor=west}},
  ]

  \addplot[
    fill=darkgray!80,
    draw=black,
    point meta=y,
    every node near coord/.style={inner ysep=5pt},
    error bars with mapped color=err,
    error bars/.cd,
        y dir=both,
        y explicit
] 
table [y error=err] {
x   y           err 
1   23766.5   305.2689672
2   44235.85714   487.5663503
3   31953.85714   347.1584752
};

  \addplot[
    fill=lightgray,
    draw=black,
    point meta=y,
    every node near coord/.style={inner ysep=5pt},
    error bars with mapped color=err,
    error bars/.cd,
        y dir=both,
        y explicit
] 
table [y error=err] {
x   y           err 
1   3161.571429   32.20065551
2   30727.42857  572.520734
3   14187.5   228.8166764
};

      \legend{\scriptsize{Unmodified ART}, \scriptsize{\TheName{}}}
  \end{axis}
\end{tikzpicture}
\caption{Performance Measured by CF-Bench~\cite{cfbench:chainfire}.}
\label{fig:pcb}
\end{figure}

As \TheName{} traces and extracts instructions at runtime, it slows the ART during instruction execution. To
learn the performance overhead introduced by \TheName{}, we use CF-Bench~\cite{cfbench:chainfire}
to compare the performance of the unmodified ART and ART with \TheName{}. For each environment, 
we run CF-Bench for $30$ times, and the results are presented in Figure~\ref{fig:pcb}.
A higher score indicates a better performance.
It shows that
\TheName{} brings $7.5$x, $1.4$x, $2.3$x overhead on Java score, native score, and overall score,
respectively.

Moreover, we evaluate the launch time of three popular applications (i.e., Snapchat,
Instagram, and WhatsApp) downloaded from Google Play.
While an activity in an application is launching, the \texttt{ActivityManager} reports the time usage for 
initializing and displaying. We launch each application for $30$ times and the result is summarized
in Table~\ref{table:tcdv}. The result shows that \TheName{} introduces about two times slowdown
on the launch time, and this result matches the overall overhead tested by CF-Bench.

Since our system is designed for security analyst instead of traditional users, we do not 
take performance as a critical factor. In summary, we consider the overhead is
acceptable and leave the further improvement as our future work.

\section{Related Work}
\label{sec:related}

\subsection{Static Analysis Tools}

FlowDroid~\cite{flowdroid:pldi14} is a static taint-analysis tool for Android
applications,
and it achieves a high accuracy by mitigating the gaps between lifecycle methods and
callback methods. Amandroid~\cite{amandroid:ccs14} and IccTA~\cite{iccta:icse15} 
aim to resolve the implicit control flows during inter-component communication.
EdgeMiner~\cite{edgeminer:ndss15} 
links the callback methods with their registration methods
to facilitate the static analysis tools in gaining more precise results. DroidSafe~\cite{droidsafe:ndss15} 
implements a simplified model of 
the Android system and solves native code in the Android framework by manually analyzing 
the source code and writing stubs for them in Java. HornDroid~\cite{horndroid:eurosp16} 
generates Horn clauses from the bytecode of application 
and performs both value-sensitive and flow-sensitive analysis on the clauses. 
HSOMiner~\cite{hsominer:ndss17} uses machine learning algorithms to discover the
hidden sensitive operations by analyzing the branch instructions and their related
conditional branches.

\setlength{\textfloatsep}{5pt}

\setlength\tabcolsep{7pt}
\begin{table}
\centering
\caption{Time Consumption of \TheName{}. 
The column "Original" 
represents the mean and standard deviation (STD) of the launch time with unmodified ART,
while the last column represents launch time with \TheName{}.}
\scriptsize
\begin{tabular}{lcrrrr} \\
\toprule
  \multicolumn{1}{ l }{\multirow{2}{*}{Application}}
  & \multicolumn{1}{ c }{\multirow{2}{*}{Version}}
  & \multicolumn{2}{ c }{Original}
  & \multicolumn{2}{ c }{With \TheName{}} \\
\cmidrule(r){3-4}
\cmidrule(r){5-6}
  & & Mean & STD & Mean & STD \\
\midrule
  \multicolumn{1}{ l }{Snapchat} & 
  \multicolumn{1}{ l }{9.43.0.0} & 
  \multicolumn{1}{ r }{826.9ms} & 
  \multicolumn{1}{ r }{52.11ms} &
  \multicolumn{1}{ r }{1,664.7ms} &
  \multicolumn{1}{ r }{16.08ms} \\
  \multicolumn{1}{ l }{Instagram} & 
  \multicolumn{1}{ l }{9.7.0} & 
  \multicolumn{1}{ r }{608.5ms} & 
  \multicolumn{1}{ r }{45.6ms} & 
  \multicolumn{1}{ r }{1,275.8ms} & 
  \multicolumn{1}{ r }{25.37ms}\\
  \multicolumn{1}{ l }{WhatsApp} & 
  \multicolumn{1}{ l }{2.16.310} &
  \multicolumn{1}{ r }{236.4ms} &
  \multicolumn{1}{ r }{12.24ms}  &
  \multicolumn{1}{ r }{480.2ms} &
  \multicolumn{1}{ r }{84.3ms} \\
\bottomrule
\end{tabular}
\label{table:tcdv}
\end{table}

\subsection{Dynamic Analysis Tools}

DroidScope~\cite{droidscope:usenix12} provides an instrumentation tool to monitor
the executed bytecode and native instructions to help analysts learn the malware manually.
VetDroid~\cite{vetdroid:ccs13} executes the Android applications by a custom application
driver and performs a permission usage behavior analysis. CopperDroid~\cite{copperdroid:ndss15}
traces the system calls and reconstructs the behavior of the target application. 
TaintDroid~\cite{taintdroid:osdi10} and TaintART~\cite{taintart:ccs16} are taint flow analysis
system on different Android Java virtual machines. They track the information flow of the
target application at runtime and report the data leakage from sink methods. 
DexHunter~\cite{dexhunter:esorics15} focuses on how 
to dump the whole DEX file from memory at a \enquote{right timing}. 
AppSpear~\cite{appspear:raid15} leverages the key data structures in Dalvik to reassemble 
the DEX file and claims that these data structures are reliable. 
Both DexHunter and AppSpear assume that there is a clear boundary between the unpacking code
and the original code. However, the unpacking code and malicious code may intersperse with each other. 
Moreover, advanced malware can modify bytecode and data in the DEX file at runtime, 
and thus the previous dump-based
unpacking systems will miss the content modified after the dump procedure.

\subsection{Hybrid Analysis Tools}

Harvester~\cite{harvester:ndss16} collects runtime values and injects these values
into the DEX file for the accuracy improvement of analysis tools.
However, some limitations still exist. Firstly, marking logging points and backward slicing are based on
the original DEX file. If packing is considered, Harvester loses its target like other static analysis tools.
In contrast, \TheName{} does not analyze the original DEX file. 
Additionally, Harvester greatly facilitates static analysis tools on solving reflections as they reduce the parameters
back into constant strings. However, malware can use advanced reflection code
to evade the analysis. Since \TheName{} replaces the reflective call
with direct call, we do not care about how the adversaries use reflection. 

\subsection{Unpacking and Reassembling in Traditional Platforms}

Ugarte et al.~\cite{sok:sp15} present a summary of recent unpacking tools and develop
an analysis framework for measuring the complexity of a large variety of packers. 
CoDisasm~\cite{codisasm:ccs15} is a dissembler tool that takes memory snapshot during execution and
disassembles the captured memory. Uroboros~\cite{uroboros:usenix15} aims to disassemble binaries with
a reassembleable approach. Their reassembling method is based on the disassembling output of Uroboros. 
\TheName{} is different from these systems as we do not disassemble the binary or monitor memory. 
~\cite{yadegari:sp15} collects the instruction trace at runtime and performs taint analysis
on the trace. Unlike ~\cite{yadegari:sp15}, \TheName{} aims to facilitate the other static analysis tools and
outputs a standardized DEX file, which could be used for state-of-the-art static analysis tools to perform
different kinds of analysis including taint analysis.

\section{Limitations and Future Work}
\label{sec:discussion}


Although the bytecode collection in \TheName{} is not based on the machine code in ART, 
the experience of TaintART shows that we can also implement our collecting algorithm in the 
compliers of ART~\cite{taintart:ccs16} to achieve the same goal. As we implement \TheName{} in a real mobile device, 
we consider that it is transparent to 
applications with anti-emulation techniques. However, 
advanced malware may be aware of its
existence by code footprints or checksum values of Android libraries. 
One potential solution is to leverage hardware isolated execution environment mentioned 
in~\cite{ninja:usenixsecurity17} to reduce the artifacts of the system and improve the 
transparency.
The code coverage improvement modules in \TheName{}
may introduce additional false positives on the 
unreachable code paths caused by unrealistic input. It is a trade-off between the code coverage
and the analysis precision.
As \TheName{} collects instructions in ART, our procedure may also be compromised by native
code. To prevent attackers tampering \TheName{}, we can randomize the memory address 
of \TheName{}~\cite{morula:sp14, blender:raid16} to make it difficult to be located. 
Additionally, 
using 
sandbox~\cite{goingnative:ndss16,nativeguard:wisec14} or
hardware-assisted isolated execution environments such as TrustZone 
technology~\cite{trustshadow:mobisys17,ninja:usenixsecurity17,sok:hasp16,case:sp16}
can secure the execution of \TheName{}. Note that applying these techniques to the entire ART may
introduce a heavy performance overhead or compatibility issues, and we need to restrictively use them
on \TheName{} only.
Currently, \TheName{} only reveals the behavior performed by Java code. However, JNI
technique allows sophisticated malware to perform malicious behavior through native code.
We consider tracing the native instructions and reassemble them as our future work.
\section{Conclusions}
\label{sec:conclusions}

In this paper, we present \TheName{}, a novel system that performs bytecode extraction
and reassembling for aiding static analysis.
It adopts instruction-level JIT collection to record the data and control flows 
of applications, and reassembles the extracted information back into a
new DEX file. 
The evaluation results on packed DroidBench samples and real-world applications 
with state-of-the-art static analysis tools show that  
\TheName{} correctly reveals the behavior in packed applications even with 
self-modifying code. The F-Measures of FlowDroid, DroidSafe, and HornDroid 
increase by $33.3\%$, $31.1\%$, and $23.6\%$, respectively. We also
show that \TheName{} provides a better accuracy than pure dynamic analysis,
and our force execution module efficiently increases the code coverage of
the dynamic analysis.
\section{Acknowledgement}
\label{sec:ack}

This work is supported by the National Science Foundation Grant No. CICI-1738929 and IIS-1724227. Opinions, findings, conclusions and recommendations expressed in this material are those of the authors and do not necessarily reflect the views of the US Government.

\balance
\bibliographystyle{IEEEtran}
\bibliography{IEEEabrv,bibliography/fengwei}

\begin{thebibliography}{10}
\providecommand{\url}[1]{#1}
\csname url@samestyle\endcsname
\providecommand{\newblock}{\relax}
\providecommand{\bibinfo}[2]{#2}
\providecommand{\BIBentrySTDinterwordspacing}{\spaceskip=0pt\relax}
\providecommand{\BIBentryALTinterwordstretchfactor}{4}
\providecommand{\BIBentryALTinterwordspacing}{\spaceskip=\fontdimen2\font plus
\BIBentryALTinterwordstretchfactor\fontdimen3\font minus
  \fontdimen4\font\relax}
\providecommand{\BIBforeignlanguage}[2]{{%
\expandafter\ifx\csname l@#1\endcsname\relax
\typeout{** WARNING: IEEEtran.bst: No hyphenation pattern has been}%
\typeout{** loaded for the language `#1'. Using the pattern for}%
\typeout{** the default language instead.}%
\else
\language=\csname l@#1\endcsname
\fi
#2}}
\providecommand{\BIBdecl}{\relax}
\BIBdecl

\bibitem{flowdroid:pldi14}
S.~Arzt, S.~Rasthofer, C.~Fritz, E.~Bodden, A.~Bartel, J.~Klein, Y.~Le~Traon,
  D.~Octeau, and P.~McDaniel, ``{FlowDroid: Precise context, flow, field,
  object-sensitive and lifecycle-aware taint analysis for Android apps},'' in
  \emph{Proceedings of the 35th ACM SIGPLAN Conference on Programming Language
  Design and Implementation (PLDI'14)}, 2014.

\bibitem{horndroid:eurosp16}
S.~Calzavara, I.~Grishchenko, and M.~Maffei, ``{HornDroid: Practical and sound
  static analysis of Android applications by SMT solving},'' in
  \emph{Proceedings of the 1st IEEE European Symposium on Security and Privacy
  (EuroS\&P'16)}, 2016.

\bibitem{droidsafe:ndss15}
M.~I. Gordon, D.~Kim, J.~H. Perkins, L.~Gilham, N.~Nguyen, and M.~C. Rinard,
  ``{Information flow analysis of Android applications in DroidSafe},'' in
  \emph{Proceedings of the 22nd Network and Distributed System Security
  Symposium (NDSS'15)}, 2015.

\bibitem{iccta:icse15}
L.~Li, A.~Bartel, T.~F. Bissyand{\'e}, J.~Klein, Y.~Le~Traon, S.~Arzt,
  S.~Rasthofer, E.~Bodden, D.~Octeau, and P.~McDaniel, ``{IccTA: Detecting
  inter-component privacy leaks in Android apps},'' in \emph{Proceedings of the
  37th International Conference on Software Engineering-Volume 1 (ICSE'15)},
  2015.

\bibitem{amandroid:ccs14}
F.~Wei, S.~Roy, X.~Ou, and Robby, ``{Amandroid: A precise and general
  inter-component data flow analysis framework for security vetting of Android
  apps},'' in \emph{Proceedings of the 21st ACM SIGSAC Conference on Computer
  and Communications Security (CCS'14)}, 2014.

\bibitem{report:avl15}
{AVL Team}, ``{AVL malware report 2015},'' \url{https://www.avlsec.com/}, 2016.

\bibitem{ali:protector}
{Alibaba Inc.}, ``{AliProtector},'' \url{http://jaq.alibaba.com/}, 2014.

\bibitem{baidu:protector}
{Baidu Inc.}, ``{BaiduProtector},'' \url{http://app.baidu.com/jiagu/}, 2014.

\bibitem{ijiami:protector}
{Ijiami Inc.}, ``{IJiamiProtector},'' \url{http://www.ijiami.cn/AppProtect},
  2013.

\bibitem{dex:protector}
{Licel Inc.}, ``{DexProtector},'' \url{https://dexprotector.com/}, 2013.

\bibitem{360:protector}
{Qihoo 360 Inc.}, ``{360Protector},'' \url{http://jiagu.360.cn/protection},
  2014.

\bibitem{tencent:protector}
{Tencent Inc.}, ``{TencentProtector},'' \url{http://legu.qcloud.com/}, 2014.

\bibitem{appspear:raid15}
W.~Yang, Y.~Zhang, J.~Li, J.~Shu, B.~Li, W.~Hu, and D.~Gu, ``{AppSpear:
  Bytecode decrypting and DEX reassembling for packed Android malware},'' in
  \emph{Proceedings of the 18th International Symposium on Research in Attacks,
  Intrusions and Defenses (RAID'15)}, 2015.

\bibitem{dexhunter:esorics15}
Y.~Zhang, X.~Luo, and H.~Yin, ``{DexHunter: Toward extracting hidden code from
  packed Android applications},'' in \emph{Proceedings of the 20th European
  Symposium on Research in Computer Security (ESORICS'15).}, 2015.

\bibitem{asac:bb}
{Bluebox Security Inc.}, ``{Android security analysis challenge: Tampering
  Dalvik bytecode during runtime},''
  \url{https://bluebox.com/android-security-analysis-challenge-tampering-dalvik-bytecode-during-runtime/},
  2013.

\bibitem{dabid:bh15}
J.~hyuk Jung and J.~Lee, ``{DABID: The powerful interactive Android debugger
  for Android malware analysis},'' Asia Black Hat, 2015.

\bibitem{taintdroid:osdi10}
W.~Enck, P.~Gilbert, B.-G. Chun, L.~P. Cox, J.~Jung, P.~McDaniel, and A.~N.
  Sheth, ``{TaintDroid: an information-flow tracking system for realtime
  privacy monitoring on smartphones},'' in \emph{Proceedings of the 9th USENIX
  Symposium on Operating Systems Design and Implementation (OSDI'10)}, 2010.

\bibitem{taintart:ccs16}
M.~Sun, T.~Wei, and J.~Lui, ``{TaintART: a practical multi-level
  information-flow tracking system for Android RunTime},'' in \emph{Proceedings
  of the 23rd ACM SIGSAC Conference on Computer and Communications Security
  (CCS'16)}, 2016.

\bibitem{copperdroid:ndss15}
K.~Tam, S.~J. Khan, A.~Fattori, and L.~Cavallaro, ``{CopperDroid: Automatic
  reconstruction of Android malware behaviors},'' in \emph{Proceedings of the
  22nd Network and Distributed System Security Symposium (NDSS'15)}, 2015.

\bibitem{droidscope:usenix12}
L.~K. Yan and H.~Yin, ``Droidscope: seamlessly reconstructing the os and dalvik
  semantic views for dynamic android malware analysis,'' in \emph{Proceedings
  of the 21st USENIX Security Symposium (USENIX Security'12)}, 2012.

\bibitem{vetdroid:ccs13}
Y.~Zhang, M.~Yang, B.~Xu, Z.~Yang, G.~Gu, P.~Ning, X.~S. Wang, and B.~Zang,
  ``{Vetting undesirable behaviors in Android apps with permission use
  analysis},'' in \emph{Proceedings of the 20th ACM SIGSAC Conference on
  Computer and Communications Security (CCS'13)}, 2013.

\bibitem{barros:ase15}
P.~Barros, R.~Just, S.~Millstein, P.~Vines, W.~Dietl, M.~d'Amorim, and M.~D.
  Ernst, ``{Static analysis of implicit control flow: Resolving {Java}
  reflection and {Android} intents},'' in \emph{Proceedings of the 30th Annual
  International Conference on Automated Software Engineering (ASE'15)}, 2015.

\bibitem{harvester:ndss16}
S.~Rasthofer, S.~Arzt, M.~Miltenberger, and E.~Bodden, ``{Harvesting runtime
  values in Android applications that feature anti-analysis techniques},'' in
  \emph{Proceedings of the 23rd Network and Distributed System Security
  Symposium (NDSS'16)}, 2016.

\bibitem{droidbench:ecssse}
{EC SPRIDE Secure Software Engineering Group}, ``{DroidBench},''
  \url{https://github.com/secure-software-engineering/DroidBench}, 2013.

\bibitem{tamiflex:icse11}
E.~Bodden, A.~Sewe, J.~Sinschek, H.~Oueslati, and M.~Mezini, ``{Taming
  reflection: Aiding static analysis in the presence of reflection and custom
  class loaders},'' in \emph{Proceedings of the 33rd International Conference
  on Software Engineering}, 2011.

\bibitem{fdroid:fdroid}
{F-Droid}, ``{F-Droid},'' \url{https://f-droid.org/}, 2011.

\bibitem{sapienz:issta16}
K.~Mao, M.~Harman, and Y.~Jia, ``{Sapienz: Multi-objective automated testing
  for Android applications},'' in \emph{Proceedings of the 25th ACM SIGSOFT
  International Symposium on Software Testing and Analysis (ISSTA'16)}, 2016.

\bibitem{guiripper:ase12}
D.~Amalfitano, A.~R. Fasolino, P.~Tramontana, S.~De~Carmine, and A.~M. Memon,
  ``{Using GUI ripping for automated testing of Android applications},'' in
  \emph{Proceedings of the 27th IEEE/ACM International Conference on Automated
  Software Engineering (ASE'12)}, 2012.

\bibitem{a3e:oopsla13}
T.~Azim and I.~Neamtiu, ``{Targeted and depth-first exploration for systematic
  testing of Android apps},'' in \emph{Proceedings of the 19th ACM SIGPLAN
  International Conference on Object Oriented Programming Systems Languages \&
  Applications (OOPSLA'13)}, 2013.

\bibitem{monkey:gg}
{Google Inc.}, ``{UI/Application Exerciser Monkey},''
  \url{https://developer.android.com/studio/test/monkey.html}, 2008.

\bibitem{puma:mobisys14}
S.~Hao, B.~Liu, S.~Nath, W.~G. Halfond, and R.~Govindan, ``{Puma: Programmable
  ui-automation for large-scale dynamic analysis of mobile apps},'' in
  \emph{Proceedings of the 12th Annual International Conference on Mobile
  systems, applications, and services (MobiSys'14)}, 2014.

\bibitem{dynodroid:fse13}
A.~Machiry, R.~Tahiliani, and M.~Naik, ``{Dynodroid: An input generation system
  for Android apps},'' in \emph{Proceedings of the 9th Joint Meeting of the
  European Software Engineering Conference and the ACM SIGSOFT Symposium on the
  Foundations of Software Engineering (ESEC'13/FSE'13)}, 2013.

\bibitem{anand:fse12}
S.~Anand, M.~Naik, M.~J. Harrold, and H.~Yang, ``{Automated concolic testing of
  smartphone apps},'' in \emph{Proceedings of the 20th ACM SIGSOFT
  International Symposium on the Foundations of Software Engineering (FSE'12)},
  2012.

\bibitem{klee:osdi08}
C.~Cadar, D.~Dunbar, and D.~R. Engler, ``{KLEE: Unassisted and automatic
  generation of high-coverage tests for complex systems programs},'' in
  \emph{Proceedings of the 8th USENIX Symposium on Operating Systems Design and
  Implementation (OSDI'08)}, 2008.

\bibitem{se:sigsoft12}
N.~Mirzaei, S.~Malek, C.~S. P{\u{a}}s{\u{a}}reanu, N.~Esfahani, and R.~Mahmood,
  ``{Testing Android apps through symbolic execution},'' \emph{ACM SIGSOFT
  Software Engineering Notes}, 2012.

\bibitem{intellidroid:ndss16}
M.~Y. Wong and D.~Lie, ``{IntelliDroid: A targeted input generator for the
  dynamic analysis of Android malware},'' in \emph{Proceedings of the 23nd
  Network and Distributed System Security Symposium (NDSS'16)}, 2016.

\bibitem{appintent:ccs13}
Z.~Yang, M.~Yang, Y.~Zhang, G.~Gu, P.~Ning, and X.~S. Wang, ``{Appintent:
  analyzing sensitive data transmission in Android for privacy leakage
  detection},'' in \emph{Proceedings of the 20th ACM SIGSAC Conference on
  Computer and Communications Security (CCS'13)}, 2013.

\bibitem{iris:ccs15}
Z.~Deng, B.~Saltaformaggio, X.~Zhang, and D.~Xu, ``{iRiS: Vetting private api
  abuse in iOS applications},'' in \emph{Proceedings of the 22nd ACM SIGSAC
  Conference on Computer and Communications Security (CCS'15)}, 2015.

\bibitem{jforce:www17}
K.~Kim, I.~L. Kim, C.~H. Kim, Y.~Kwon, Y.~Zheng, X.~Zhang, and D.~Xu,
  ``{J-Force: Forced Execution on JavaScript},'' in \emph{Proceedings of the
  26th International Conference on World Wide Web (WWW'17)}, 2017.

\bibitem{xforce:usenix14}
F.~Peng, Z.~Deng, X.~Zhang, D.~Xu, Z.~Lin, and Z.~Su, ``{X-Force:
  Force-executing binary programs for security applications},'' in
  \emph{Proceedings of the 23rd USENIX Security Symposium (USENIX
  Security'14)}, 2014.

\bibitem{asop:gg}
{Google Inc.}, ``{Android open source project},''
  \url{https://source.android.com/}, 2008.

\bibitem{twrp:teamwin}
{Team Win}, ``{Team win recovery project},'' \url{https://twrp.me/}, 2014.

\bibitem{dex:gg}
{Google Inc.}, ``{Dalvik executable format},''
  \url{https://source.android.com/devices/tech/dalvik/dex-format.html}, 2008.

\bibitem{soot:cetus11}
P.~Lam, E.~Bodden, O.~Lhot{\'a}k, and L.~Hendren, ``{The Soot framework for
  Java program analysis: A retrospective},'' in \emph{Proceedings of the Cetus
  Users and Compiler Infastructure Workshop (CETUS'11)}, 2011.

\bibitem{bangcle:protector}
{Bangcle Ltd.}, ``{BangcleProtector},'' \url{https://www.bangcle.com/}, 2013.

\bibitem{netqin:protector}
{NQ Mobile}, ``{NetQinProtector},'' \url{https://shield.nq.com}, 2014.

\bibitem{apkprotect:protector}
{Android APK Encryption and Protection}, ``{APKProtector},''
  \url{https://sourceforge.net/projects/apkprotect/}, 2013.

\bibitem{playdrone:sigmetrics14}
N.~Viennot, E.~Garcia, and J.~Nieh, ``{A measurement study of Google Play},''
  in \emph{Proceedings of the ACM SIGMETRICS}, 2014.

\bibitem{googleplay:gg}
{Google}, ``{Google Play},'' \url{https://play.google.com/store?hl=en}, 2017.

\bibitem{360market:360}
{Qihoo 360 Inc.}, ``{360 Market},'' \url{http://zhushou.360.cn/}, 2017.

\bibitem{wandoujia:wandoujia}
{Wandoujia}, ``{Wandoujia Market},'' \url{http://www.wandoujia.com/apps}, 2017.

\bibitem{fdroidrandom:fdroid}
{F-Droid}, ``{F-Droid Random Page},''
  \url{https://f-droid.org/wiki/page/Special:Random}, 2011.

\bibitem{jacoco:jacoco}
{JaCoCo}, ``{Java Code Coverage Library},''
  \url{http://www.eclemma.org/jacoco/}, 2009.

\bibitem{cfbench:chainfire}
Chainfire, ``{CF-Bench},''
  \url{https://play.google.com/store/apps/details?id=eu.chainfire.cfbench},
  2013.

\bibitem{edgeminer:ndss15}
Y.~Cao, Y.~Fratantonio, A.~Bianchi, M.~Egele, C.~Kruegel, G.~Vigna, and
  Y.~Chen, ``{EdgeMiner: Automatically detecting implicit control flow
  transitions through the Android framework},'' in \emph{Proceedings of the
  22nd Network and Distributed System Security Symposium (NDSS'15)}, 2015.

\bibitem{hsominer:ndss17}
X.~Pan, X.~Wang, Y.~Duan, X.~Wang, and H.~Yin, ``{Dark Hazard: Learning-based,
  large-scale discovery of hidden sensitive operations in Android apps},'' in
  \emph{Proceedings of the 24th Network and Distributed System Security
  Symposium (NDSS'17)}, 2017.

\bibitem{sok:sp15}
X.~Ugarte-Pedrero, D.~Balzarotti, I.~Santos, and P.~G. Bringas, ``{SoK: Deep
  packer inspection: A longitudinal study of the complexity of run-time
  packers},'' in \emph{Proceedings of the 36th IEEE Symposium on Security and
  Privacy (S\&P'15)}, 2015.

\bibitem{codisasm:ccs15}
G.~Bonfante, J.~Fernandez, J.-Y. Marion, B.~Rouxel, F.~Sabatier, and
  A.~Thierry, ``{CoDisasm: Medium scale concatic disassembly of self-modifying
  binaries with overlapping instructions},'' in \emph{Proceedings of the 22nd
  ACM SIGSAC Conference on Computer and Communications Security (CCS'15)},
  2015.

\bibitem{uroboros:usenix15}
S.~Wang, P.~Wang, and D.~Wu, ``{Reassembleable disassembling},'' in
  \emph{Proceedings of the 24th USENIX Security Symposium (USENIX
  Security'15)}, 2015.

\bibitem{yadegari:sp15}
B.~Yadegari, B.~Johannesmeyer, B.~Whitely, and S.~Debray, ``{A generic approach
  to automatic deobfuscation of executable code},'' in \emph{Proceedings of the
  36th IEEE Symposium on Security and Privacy (S\&P'15)}, 2015.

\bibitem{ninja:usenixsecurity17}
\BIBentryALTinterwordspacing
Z.~Ning and F.~Zhang, ``Ninja: Towards transparent tracing and debugging on
  arm,'' in \emph{26th USENIX Security Symposium (USENIX Security 17)}.\hskip
  1em plus 0.5em minus 0.4em\relax Vancouver, BC: USENIX Association, 2017.
  [Online]. Available:
  \url{https://www.usenix.org/conference/usenixsecurity17/technical-sessions/presentation/ning}
\BIBentrySTDinterwordspacing

\bibitem{morula:sp14}
B.~Lee, L.~Lu, T.~Wang, T.~Kim, and W.~Lee, ``{From zygote to morula:
  Fortifying weakened ASLR on Android},'' in \emph{Proceedings of the 35th IEEE
  Symposium on Security and Privacy (S\&P'14)}, 2014.

\bibitem{blender:raid16}
M.~Sun, J.~C. Lui, and Y.~Zhou, ``{Blender: Self-randomizing address space
  layout for Android apps},'' in \emph{Proceedings of the 19th International
  Symposium on Research in Attacks, Intrusions and Defenses (RAID'16)}, 2016.

\bibitem{goingnative:ndss16}
V.~Afonso, A.~Bianchi, Y.~Fratantonio, A.~Doup{\'e}, M.~Polino, P.~de~Geus,
  C.~Kruegel, and G.~Vigna, ``{Going Native: Using a large-scale analysis of
  {Android} apps to create a practical native-code sandboxing policy},'' in
  \emph{Proceedings of the 23nd Network and Distributed System Security
  Symposium (NDSS'16)}, 2016.

\bibitem{nativeguard:wisec14}
M.~Sun and G.~Tan, ``{NativeGuard: Protecting android applications from
  third-party native libraries},'' in \emph{Proceedings of the 2014 ACM
  conference on Security and privacy in wireless \& mobile networks
  (WiSec'14)}, 2014.

\bibitem{trustshadow:mobisys17}
L.~Guan, P.~Liu, X.~Xing, X.~Ge, S.~Zhang, M.~Yu, and T.~Jaeger,
  ``{TrustShadow: Secure execution of unmodified applications with ARM
  trustzone},'' in \emph{Proceedings of the 15th Annual International
  Conference on Mobile systems, applications, and services (MobiSys'17)}, 2017.

\bibitem{sok:hasp16}
F.~Zhang and H.~Zhang, ``{SoK: A study of using hardware-assisted isolated
  execution environments for security},'' in \emph{Proceedings of Hardware and
  Architectural Support for Security and Privacy (HASP'16)}, 2016.

\bibitem{case:sp16}
N.~Zhang, K.~Sun, W.~Lou, and Y.~T. Hou, ``{CaSE: Cache-Assisted Secure
  Execution on ARM Processors},'' in \emph{Proceedings of the 37th IEEE
  Symposium on Security and Privacy (S\&P'16)}, 2016.

\end{thebibliography}



%



\end{document}